\documentclass[fleqn,usenatbib]{mnras}

\usepackage{newtxtext,newtxmath}

\usepackage[T1]{fontenc}
\usepackage[english]{babel}
\usepackage{color}

\DeclareRobustCommand{\VAN}[3]{#2}
\let\VANthebibliography\thebibliography
\def\thebibliography{\DeclareRobustCommand{\VAN}[3]{##3}\VANthebibliography}

\usepackage{graphicx}	
\usepackage{amsmath}	
\usepackage{booktabs}     
\usepackage{multirow}
\usepackage{siunitx}

\title{Correlating Particle Acceleration Rates with Plasma Conditions in Colliding Wind Binaries}

\author[Cordeiro, G. B.]{
Gislaine B. Cordeiro$^{1}$\thanks{E-mail: gislaine.cordeiro@inpe.br}; Diego A. Falceta-Gonçalves$^{2}$\thanks{E-mail: dfalceta@usp.br}; Grzegorz Kowal$^{2}$, Vanessa F. Giraldez-Garcia$^{2}$
\\
$^{1}$Instituto Nacional de Pesquisas Espaciais - Av. dos Astronautas, 1.758 - Jardim da Granja, São José dos Campos - SP, 12227-010, Brazil\\
$^{2}$Escola de Artes, Ci\^encias e Humanidades, Universidade de S\~ao Paulo, Rua Arlindo Bettio 1000, CEP 03828-000,
S\~ao Paulo, Brazil}

\date{Accepted XXX. Received YYY; in original form ZZZ}

\pubyear{2025}

\begin{document}
\maketitle


\begin{abstract}
Recent observations have revealed star-forming regions as possible origin sites of very-high-energy (TeV) cosmic rays, not associated with supernova remnants. Colliding-wind binaries (CWBs) are strong X-ray and radio synchrotron emitters and have been proposed as potential accelerators of such particles. We perform high-resolution three-dimensional magnetohydrodynamic simulations coupled with test-particle integration to investigate how local plasma conditions affect particle acceleration in CWBs. We find that the maximum particle energies and the hardness of the energy distributions depend on the shock magnetization and cooling efficiency. For moderate magnetization (>1 G), CWBs can accelerate hadronic particles up to hundreds of TeV or even PeV energies, with more than 1\% of particles reaching the very-high-energy range. By correlating the local acceleration rate with plasma quantities — \textit{magnetic field strength, current density, vorticity, and velocity divergence} — we show that turbulence and magnetic field complexity dominate the acceleration, while classical diffusive shock acceleration plays a limited role. These results suggest that turbulent, magnetically driven processes are key to producing relativistic particles in CWBs, with implications for future high sensitivity $\gamma$-ray observations (e.g. LACT and CTAO).
\end{abstract}



\section{Introduction} 

SNe have long been considered as the major accelerators of high energy cosmic rays (CRs) within the Galaxy \cite{ginzburg1961}, at least for energies as high as $\sim 1$TeV. At the VHE regime ($> 100$TeV) however, the extent of the SNe contribution is still unclear. Recent data obtained with LLHASO experiment reveal extended high energy emission from star forming regions (SFRs), indicating that other sources may contribute - if not dominate - the acceleration/transport of the VHE particles detected \citep{llhaso21,cygnusx23,llhaso24}. Part of these diffuse emitting regions are clearly associated to young massive star clusters. Molecular cloud  (MC) association to the LLHASO VHE sources was also confirmed by \citet{tsuji25}. Among possible alternate engines for cosmic particle acceleration in SFRs we may cite other compact object related scenarios (such as  shocks of pulsar nebulae, magnetic reconnection and centrifugal effects in their outflows), and non-compact object related phenomena, such as shocks in  massive star clusters (MSCs) outflows \citep[e.g.][]{morlino21}, and WR/O stellar wind-wind collisions \citep{eichler1993particle,falceta15,pittard20, kowal.2021}.  

Colliding wind binaries (CWBs) have been proposed as a possible Galactic source of high energy cosmic ray particles. \citet{eichler1993particle} conjectured CWBs as significant sources of non-thermal radiation, including gamma rays, due to the acceleration of particles in the supersonic shocks formed as the stellar winds interact. That work represents the theoretical groundwork for understanding particle acceleration and the resulting non-thermal radio emission in these systems, by proposing that diffusive shock acceleration (DSA) occurs at the wind collision interface. Since then, many observational works have focused in detecting the non-thermal nature of massive binaries, mostly succeeding in detecting synchrotron radiation at radio wavelengths \cite[e.g.][]{white1995,skinner1999,monnier2002,cappa2004,doug2005,abraham2005,becker2007,falceta2012,de2013catalogue,becker2017,benaglia2020,saha2023,benaglia2025}. 

Contributions of CWBs to the high energy end of the electromagnetic spectrum occur both in X-rays and gamma-rays. In the X-rays, most of the emission comes from thermal free-free emission, produced by the hot plasma confined at the shock region \citep[e.g.][]{stevens1992,falceta2005,abraham2010,rauw2016,rauw2025}. In the gamma-rays, emission can occur by inverse Compton (IC) from relativistic leptons interacting with background radiation and by proton-proton collisions and further $\pi^{0}$-decay generating high energy photons. Due to current instrument sensitivities, clear detection of gamma-ray emission from CWBs has only occurred for the massive star system of $\eta$ Carinae \citep[see][]{tavani2009detection,marti2021}. Detection of X-rays from non-thermal origin was also reported from $\eta$ Carinae \citep{hamaguchi2018non}. \cite{werner2013fermi} used 24 months of data from the FERMI-LAT telescope to find evidence of gamma-ray emission from a sample of seven CWB systems: WR 11 ($\gamma^2$ Velorum), WR 70, WR 125, WR 137, WR 140, WR 146, and WR 147. None of these systems was detected at the time, but upper limits for the emission from these objects were established. WR 140 is known to present both synchrotron and non-thermal X-ray emission \citep{de2013catalogue}, though gamma-rays has not been detected yet. Possible causes of the weak emission could be its orbital geometry, the power of its winds, and/or the intensity of the magnetic fields, which modify the acceleration efficiency of particles in these scenarios \cite{pittard2006radio,falceta15}. \cite{marti2020} reported periodic variations in WR 11's gamma-ray light curve, which could indicate that its gamma-ray emission is correlated with its orbital phases, making it (possibly) the second CWB detected in gamma-rays. 

These results indicate that, although many CWBs show signs of particle acceleration, the detection of their gamma-ray emission depends not only on the intrinsic physics of the systems but also on observational and local limitations. Possible future instruments with greater sensitivity, as the Cherenkov Telescope Array Observatory (CTAO), will be fundamental to confirm or rule out gamma-ray emission in other CWBs.


Particle acceleration in colliding wind binaries (CWBs) occurs in regions where plasma is highly compressed, such as shock fronts and highly turbulent zones. The dynamic structure of these systems creates favorable conditions for various acceleration mechanisms, where intense magnetic fields, supersonic velocities, and kinematic instabilities act together. One of the major open questions is identifying which of these mechanisms dominates the production of relativistic particles up to energies on the order of TeV or PeV, as suggested by observations in systems like $\eta$ Carinae and $\gamma^2$ Velorum.

It's worth noting that, according to the Hillas confinement criterion, the maximum energy ($E_{\rm max}$) a particle of charge $q$ can attain in an astrophysical environment depends on the magnetic field strength ($B$) and the size of the acceleration region ($L$) as $E_{\rm max}/c \simeq qBL$, i.e. the energy of a particle with a Larmor radius larger than the characteristic lengthscale of the system. The magnetic field is, in general, poorly known and therefore the maximum energy of particles. According to this criterium, a CWB shock region of $L \sim 1$AU, can produce TeV protons if $B > 100$mG. 

The primary mechanisms responsible for particle acceleration include:

{\it 1) Diffusive Shock Acceleration (DSA)} -- The most widely studied mechanism in shock regions is the Diffusive Shock Acceleration (DSA), which is a First-Order Fermi type acceleration, in which particles repeatedly cross the shock front being reflected by incoming magnetic field on both sides of the discontinuity \citep{blandford1978,bell1978}. For an adiabatic strong shock, with a downstream/upstream density compression factor of 4,
the energy gain at each crossing is  $\langle \frac{\Delta E}{E} \rangle \sim \frac{4}{3}\frac{U}{c}$, being $U$ the relative shock speed. Such an acceleration produces a power-law energy distribution as $N(E) \propto E^{-2}$, for stationary strong shocks. In this mechanism, though the maximum energies reached depend on magnetization level of the shock (see Hillas criterium above), the acceleration rate is not dependent on the strength of the magnetic field $B$, as it does depend on the kinematics of shock. 

{\it 2) Turbulent - Stochastic Acceleration} -- also known as 2nd Order Fermi Acceleration. In highly turbulent environments, such as those produced by the interaction of massive stellar winds, the magnetic field can be significantly distorted, with generation of MHD waves. In this mechanism, particles gain energy by randomly interacting with randonmly moving magnetic structures (Alfvén waves, turbulent vortices, magnetized "clouds"), without a preferred direction \citep{fermi1949}. These interactions lead to both energy gain and loss, depending on the trajectory of the particle, with gain events being statistically more probable than losses, resulting in an average energy gain that scales with the square of the fluctuations of velocity, $\langle \Delta E \rangle \sim (\delta v/c)^2$. Such process operates until particles escape the accelerating region (at timescale $\tau_\mathrm{esc}$), and the steady state solution of the diffusion-loss Fokker-Planck equation is $N(E) \propto E^{-(1+\frac{1}{ \alpha \tau_\mathrm{esc}})}$, where $\alpha = \frac{4}{3}\frac{\delta v^2}{cL}$ and $L$ the characteristic lengthscale of the fluctuations. 
This process is less efficient compared to DSA in weak turbulence, but potentially dominant in regions with strong disordered magnetic field geometry, where spatial diffusion is enhanced \citep{wiedemann2015particle, caprioli2024particle}. For the scattering of particles with Alfvén waves, the energy gain rate is $\frac{dE}{dt} \sim \frac{Ev_A^2}{D_{\parallel}}$ , where $D_{\parallel}$ is the parallel diffusion coefficient and $v_A = B/\sqrt{4 \pi \rho}$ the Alfvén speed \citep{jokipii1977}. Therefore, for stochastic acceleration by Alfvén waves one should expect the acceleration rate to be dependent on the magnetic field strength.

{\it 3) Magnetic Mirror Effect} -- Another relevant process is the Mirror Effect, associated with particle trapping in regions where the magnetic field geometry is complex/turbulent and compressed magnetic islands are present. The resulting magnetic gradient leads to particle reflection/acceleration, causing them to bounce between magnetic mirrors, while gaining energy in the perpendicular component. This process resembles the second order Fermi mechanism if particle energy gain is small in each island, while it can be particularly effective when combined with magnetic field compression, as occurs in oblique shocks \citep{pino2009role, kowal2012particle}, when it behaves more like a first order process.

{\it 4) Turbulent Magnetic Reconnection} -- Magnetic reconnection may provide particle energy gain through different processes.
In turbulent magnetic reconnection, as described by \citet{Lazarian1999}, the reconnection layer becomes a stochastic, volumetric region, filled with interacting magnetic flux tubes that move toward each other at the reconnection speed, $V_\mathrm{rec}$, typically a significant fraction of the local Alfvén speed. Charged particles trapped within these contracting flux tubes undergo head-on scatterings each time they reflect from a converging magnetic structure, gaining energy. Since particles, after each acceleration event, encounter magnetic field irregularities produced by the surrounding turbulence, they can be scattered back into the converging region and undergo multiple successive interactions, during which they gain a significant amount of energy \citep[see][]{kowal2012particle}. In this process, the energy gain per interaction can be expressed as $\Delta E / E \sim 2 V_{\rm rec}/c$ \citep{dalpino2005}.
Finally, magnetic reconnection also generates local electric fields through resistive effects. 
Direct resistive acceleration by parallel electric fields ($\mathbf{E}_\parallel = \eta \mathbf{J}$) \citep{jardine1996} is however expected to be ineffective in astrophysical plasmas due to the extremely low resistivity in these scenarios, making the Fermi process possibly the dominant energy gain channel.

{\it 5) Betatron Effect} -- When a charged particle moves in a region where the magnetic field strength varies along its trajectory, the magnetic moment $\mu = m v_\perp^2 / (2B)$ is approximately conserved in the adiabatic limit. 
As a consequence, an increase in \(B\) leads to a proportional increase in the perpendicular kinetic energy, with $\frac{dE_\perp}{E_\perp} \sim \frac{dB}{B}$, a process known as \emph{betatron acceleration} \citep{kulsrud2005}. 
In this mechanism, the acceleration rate depends directly on the temporal rate of change of the magnetic field intensity experienced by the particle. 
In a highly turbulent magnetic field, however, the stochastic nature of $B$ variations can lead to alternating phases of energy gain and loss, reducing the net efficiency of the process. 
Nonetheless, betatron acceleration can still provide significant \emph{one-shot} energy boosts that pre-accelerate particles, making them more susceptible to further energization by other mechanisms.

Despite the relative wealth of data of non-thermal emission detection from CWBs, the processes that originate these relativistic population of particles are still not fully understood. In this work, we seek to elucidate this problem by analyzing correlations between the particle acceleration rate and relevant physical quantities, aiming to identify the processes that most influence acceleration efficiency. Recent MHD simulations by \citet{kowal.2021} have shown that colliding-wind regions can lead to highly turbulent strong magnetic fields, and to efficient particle acceleration -- beyond TeV energies.
However, in that study, the authors did not attempt to disentangle the multiple acceleration processes contributing to the final distribution of particle energies; that is, although they showed that very high-energy gamma rays could be produced in CWBs, they did not correlate the acceleration rates with the associated physical processes.

In this work, using high-resolution MHD numerical simulations coupled with a particle trajectory integration technique, we extend the study presented by \citet{kowal.2021} to establish correlations between plasma physical parameters, velocity and magnetic field geometries, and the resulting acceleration rates. The MHD numerical modeling and particle trajectory integration method are described in Section 2, where we also present convergence tests. In Section 3, we present the main results, followed by the discussion in Section 4, and we conclude the manuscript in Section 5.

\section{Numerical Modeling}
\label{sec:modeling}

\subsection{Magnetohydrodynamic Model of Colliding-Wind Binary}
\label{sec:simulation}

Since the only known CWB with confirmed gamma ray emission is $\eta$ Carinae, this object represents a perfect case to study particle acceleration by means of numerical simulations. The setup chosen corresponds to the periastron passage of the system, and the wind parameters to those of \citet{falceta2012}.

We model a binary system consisting of primary and secondary stars placed at a distance of approximately 2~c.u., corresponding to 2.21~AU, similar to the one developed by \citet{kowal.2021}. The stars cannot be fully resolved, so  boundary conditions for their winds are set numerically by spherical regions of a radius of about 0.12~AU at which the density, pressure, outflow velocity, and magnetic field are maintained during the whole simulation. The wind density at the mask surface of the primary star ($p$) is set to $\rho_{p}=10^7$ g cm$^{-3}$, while for the secondary star ($s$) it is $100$ times smaller. We assumed the wind speeds of the primary and secondary stars to be 500 km s$^{-1}$ and 2500 km s$^{-1}$, respectively. These parameters correspond to mass loss rates of $\dot{M}_p \sim 2 \times 10^{-4}$M$_\odot$ yr$^{-1}$ and $\dot{M}_s \sim 10^{-5}$M$_\odot$ yr$^{-1}$. The wind thermal pressure at the star surfaces is determined by the formula
\begin{equation}
    P_{\rm w,n}= \left ( \frac{v_{\rm w,n}}{M_{\rm w,n}} \right )^2 \frac{\rho_{\rm w,n}}{\Gamma},
\end{equation}
where ${M}_{\rm w,n}$ is the sonic Mach number of the wind at the surface, $n = p,s$ for the primary and secondary stars, respectively, and $\Gamma_{\rm eff}$ is the effective adiabatic coefficient. We set ${M}_{\rm w,p} = 3$ and ${M}_{\rm w,s} = 15$ for the primary and secondary stars, respectively. In our model, we performed four simulations with different values of $\Gamma_{\rm eff}$. The effective adiabatic coefficient corresponds to an approximate polytropic solution for the evolution of the thermal energy density considering cooling effects, such as radiative losses, ranging from completely isothermal ($\Gamma_{\rm eff} = 1$) to fully adiabatic ($\Gamma_{\rm eff} = 5/3$). Dense strong shocks are radiative and effective adiabatic coefficients closer to unity are more suited. 

Although the stellar winds are magnetized, at the locii of the mask boundary conditions they are highly super-Alfvénic, meaning the kinetic pressure dominates over the magnetic pressure. Consequently, magnetic field lines near the stars follow the radial velocity profile, leading to a radial configuration anchored at fixed surface foot points. These foot points are numerically set in terms of an internal stellar magnetic field $B_\star$, which is a free parameter. Because of the strong wind, the magnetic field decays radially. 

To further improve convergence, we initialized the system using the ram pressure balance to define the separation surface between the regions dominated by each stellar wind. This adjustment significantly reduced the time required to reach the quasi-steady-state turbulent regime. The interstellar plasma is rapidly swept out of the computational domain during the wind expansion phase; therefore, the physical conditions analyzed in this study depend solely on the stellar wind parameters and not on any assumptions about the ambient medium.

The simulation domain spans $X \in [-6, 2]$, $Y,Z \in [-8, 8]$, with physical dimensions $8L \times 16L \times 16L$ (assuming $L \approx 1.155$~AU), corresponding to the X, Y, and Z directions, respectively. The base grid resolution was $60 \times 120 \times 120$ with Adaptive Mesh Refinement (AMR) applied in blocks of $60 \times 60 \times 60$ cells. The maximum refinement level was set to 5, resulting in an effective resolution of $960 \times 1920 \times 1920$. This corresponds to a spatial resolution of $h \approx 0.0083 \, \mathrm{c.u.} \approx 0.01 \, \mathrm{AU}$ along all directions. The refinement criterion was based on the normalized values of vorticity and current density to accurately follow shocks, turbulence, and magnetic reconnection structures. We imposed outflow boundary conditions along all directions.

The above setup is evolved by the ideal adiabatic compressible MHD system of equations:
\begin{equation}
\frac{\partial \rho}{\partial t} + \nabla \cdot \left( \rho \mathbf{u} \right) = 0,
\label{eq:mhd_1}
\end{equation}
\begin{equation}
\frac{\partial \left( \rho \mathbf{u} \right)}{\partial t} + \nabla \cdot \left[ \rho \mathbf{u} \mathbf{u} + \left( p + \frac{B^2}{2} \right) I - \mathbf{B} \mathbf{B} \right] = 0,
\label{eq:mhd_2}
\end{equation}
\begin{equation}
\frac{\partial \mathbf{B}}{\partial t} - \nabla \times \left( \mathbf{u} \times \mathbf{B} \right) = 0, \, \nabla \cdot \mathbf{B} = 0,
\label{eq:mhd_3}
\end{equation}
\begin{equation}
\frac{\partial E}{\partial t} + \nabla \cdot \left[ \left( E + p \right) \mathbf{u} - \left( \mathbf{u} \cdot \mathbf{B} \right) \mathbf{B} \right] = 0,
\label{eq:mhd_4}
\end{equation}
where $\rho$, $p$, $\mathbf{u}$, and $\mathbf{B}$ are the plasma density, thermal pressure, velocity, and magnetic field, respectively, $E = (\rho u^2/2) + p/(\Gamma_{\rm eff} - 1) + (B^2/2)$ is the total energy, and $I$ is the identity matrix in a 3D rectangular domain described on the Cartesian coordinate system.

The MHD equations were solved numerically using a high-order shock-capturing adaptive refinement Godunov-type code {\tt AMUN}\footnote{AMUN code is open source and freely available from \href{https://gitlab.com/gkowal/amun-code}{https://gitlab.com/gkowal/amun-code}.}. We employed a Gaussian Process-based reconstruction scheme with a 5-point stencil in each spatial direction and a kernel width parameter $\sigma = 9.0$, enabling accurate and smooth recovery of Riemann states \citep{2016arXiv161108084R, 2019PhDT........51R}. For the Riemann solver, we adopted the HLLD approximate flux method \citep{mignone2007simple}. Temporal integration was carried out using the 3$^{\rm rd}$-order 4-stage embedded SSP3(2)4 Strong Stability Preserving Runge-Kutta method \citep{gottlieb2011strong}, which allows simultaneous control of numerical stability and local truncation error. Both the absolute and relative tolerances were set to $10^{-3}$ to guide the adaptive time stepping during the evolution.

The system was evolved until $t = 11 \, \mathrm{c.u.}$ (corresponding to roughly 22 days considering the assumed length and velocity units), at which the interaction surface between the colliding winds stopped moving through the domain, and a turbulent region produced by their interaction became statistically stationary. Once such a state is reached, the system can be evolved for an arbitrary period of time maintaining its statistical properties.

\subsection{Test Particle Integration}
\label{sec:particles}

In order to study the behavior of test particles in the binary system evolved as described in the previous subsection, we took the last available domain snapshot (velocity and magnetic field spatial distributions) and injected $10^5$ (if not specified otherwise) ions in random locations near the turbulent region created by colliding winds with random orientations of velocity and the initial particle velocity distribution corresponding to the thermal distribution at temperature $T_0 = 10^6 \, \mathrm{K}$.

The code solves trajectory evolution of any charged particle, representing various ion species from hydrogen to iron: H$^+$, He$^+$, C$^+$, Si$^+$, Ca$^+$, and Fe$^+$. The computing time is dependent on the ion mass, therefore  for optimizing the computational time, the trajectories in the present Paper are calculated for Fe$^+$ ions. 

The trajectories of injected particles were integrated using the relativistic equation of motion for a charged particle
\begin{equation}
  \frac{d}{d t} \left( \gamma \mathbf{v} \right) = \frac{q}{m} \left( \mathbf{E} + \mathbf{v} \times \mathbf{B} \right), \quad \frac{d \mathbf{r}}{dt} = \mathbf{v}, \label{eq:motion}
\end{equation}
where $\mathbf{r}$ and $\mathbf{v}$ are the particle position and velocity, respectively, $\gamma = \left( 1 - v^2/c^2 \right)^{1/2}$ is the Lorentz factor, $q$ and $m$ are particle charge and mass, respectively, $c$ is the speed of light.

The plasma-induced electric field experienced by the particle can be fully obtained through the generalized Ohm's Law\footnote{Here, the displacement between ion and electron distributions (the Hall current) and the electron pressure gradient term were neglected, owing to the large length scale of the system under study.}: 
\begin{equation}
\mathbf{E} \simeq - \mathbf{u} \times \mathbf{B} + \eta\mathbf{j} 
\label{eq:ohm}
\end{equation}
where $\mathbf{u}$ is the plasma velocity, $\mathbf{B}$ the magnetic field, $\rho$ the gas density, and $\mathbf{j}$ the current density, all taken from the grid simulation. Although the MHD simulation is performed in the ideal limit, numerical dissipation/diffusion of the computed fields occurs at the scale of the numerical resolution, which is orders of magnitude larger than the viscosity or resistivity found in nature. For the particle integration, however, the trajectory is independent of the grid limitations, and the resistivity may be set to its physical value, computed as $\eta \simeq \frac{\nu_c m_e}{n e^2}$, where $\nu_c \sim n^2 \sigma c_s$ is the collision frequency between ions and electrons, $\sigma$ is the collision cross-section, and $c_s$ is the sound speed computed for the temperature obtained from the MHD model. For all particle integrations, however, we neglected the non-ideal terms ($\eta = 0$); therefore, the effective acceleration of particles was solely due to ideal effects.


We assumed the particle velocity $\mathbf{v}$ in Equation~(\ref{eq:motion}) to be expressed in units of the speed of light $c$ and its position in the coordinates of the grid simulation. Since the grid simulations use dimensionless units, one needs to assume the unit of plasma velocity $\mathbf{u}$ and magnetic field $\mathbf{B}$. In this work, we assumed that the plasma velocity is in units of $1000$~km s$^{-1}$, and for the magnetization level of the winds, we set the magnetic field anchored at the stellar surfaces to range from $B_{\star} = 10^{-3}$ to $10^{1}$~G. Both stars are assumed to maintain the same intensity of surface magnetic fields during the whole simulation time. This value is a model parameter and should not be confused with the real stellar magnetic strength. Due to the numerical modeling limitations, it simply represents the magnetic field which is ejected from the star together with the wind. In the particle integration, the unit of time was chosen as $48 \, \mathrm{h} = 1.728 \times 10^5$~s which, assuming the plasma velocity unit of $1000$~km s$^{-1}$, results in the unit of length equal to $L \approx 1.155 \, \mathrm{AU}$.

The integration of Equation~(\ref{eq:motion}) was done with the help of the code {\tt GAccel}\footnote{GAccel code is open source and freely available from \href{https://gitlab.com/gkowal/gaccel}{https://gitlab.com/gkowal/gaccel}.} using the 8$^{th}$ order embedded Dormand-Prince (DOP853) method \citep{hairer2008} with an adaptive time step based on an error estimator. In order to determine the values of $\mathbf{u}$ and $\mathbf{B}$ at the current particle positions, we used tri-cubic interpolation (a cubic interpolation applied dimension by dimension), enforcing the monotonicity by limiting the tangents at the interpolation nodes with the minmod TVD limiter \citep{vl1979}. In the integration of the particle trajectories, we set the absolute and relative tolerance to {\tt atol = rtol = 1e-10} and the initial time step {\tt dt = 1e-6}, if not specified otherwise.

\begin{figure}[t]
	\centering
	\includegraphics[width=0.5\textwidth]{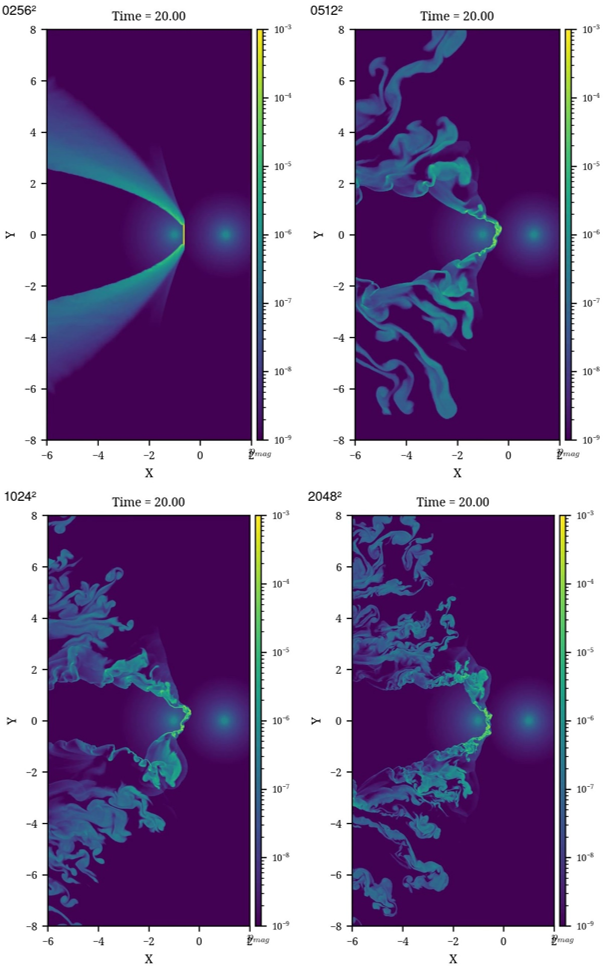}
	\caption{Convergence test of instabilities and turbulence onset in numerical simulations. Different models correspond, in sequence, to the resolutions of $256 \times 512$, $512 \times 1024$, $1024 \times 2048$, and $2048 \times 4096$ cells. Color map shows the magnetic pressure distribution of the periastron conditions of $\eta$ Carinae binary system. Length scales are in code units. }
	\label{convergence}
\end{figure}

\subsection{Convergence Test}

In numerical simulations, spatial resolution plays a crucial role in accurately capturing turbulence and the interactions between the conducting fluid and magnetic fields in shocks. At the same time, indefinite refinement of the numerical resolution is computationally prohibitive and, beyond a certain point, unnecessary. This is because the diffusive effects of waves and turbulent interactions with particle trajectories are dominated by relatively large scales, where most of the kinetic power of a typical turbulent power spectrum resides. The goal is, therefore, to determine the minimum resolution capable of developing a full turbulent inertial range spanning a few decades in length scales, allowing both diffusion processes and the natural development of Alfvén waves at the shock.

Turbulence in shocks arises from instabilities such as the Kelvin–Helmholtz instability, the Nonlinear Thin Shell Instability, the Weibel instability, and others. For turbulence to develop in the shocked plasma, the typical instability growth time must be shorter than the numerical dissipation timescale of fluctuations, which depends on the resolution. To determine the minimum grid resolution required, we performed numerical experiments, first in two dimensions, using different resolutions.

In Figure~\ref{convergence}, we present the magnetic pressure (equivalent to magnetic energy) maps of fully developed shocks obtained with the setup described above, for resolutions of $256 \times 512$, $512 \times 1024$, $1024 \times 2048$, and $2048 \times 4096$. Higher resolutions capture dynamic processes—such as turbulence development and shock compression in the wind collision region—in greater detail. We find that visible instabilities can grow within the dynamical timescales of the system at resolutions as low as $512$ cells, and that no significant visible improvement is obtained above $1024$. Therefore, our choice of $960 \times 1920 \times 1920$ cells is more than sufficient to capture the key dynamics of a three-dimensional, turbulently evolving shock, providing a good balance between numerical accuracy and computational feasibility.

\section{Results}

\subsection{Spatial Distribution of Plasma}

Before studying the distribution of test particles, we must first spatially characterize the physical properties of the wind–wind collision region in the different numerical models.

These three-dimensional scenarios provide realistic representations -- within numerical limitations -- of the physical conditions encountered by particles propagating in the shock region of $\eta$ Carinae. The strong anisotropy of the medium, the presence of strong shocks or highly magnetized regions, and the developed turbulence create a favorable environment for particle acceleration. A detailed spatial analysis of these quantities forms a solid foundation for interpreting the energy distributions and statistical correlations discussed in the following sections.

As mentioned earlier, the level of magnetization is not relevant for the magnetohydrodynamics of the plasma, given that the kinetic and thermal pressures exceed the magnetic pressure by orders of magnitude. Therefore, among the free parameters of our models, the effective adiabatic coefficient $\Gamma_{\rm eff}$ is the only one responsible for the distinct spatial distribution of the plasma’s physical properties.

\begin{figure*}
	\centering
	\includegraphics[width=1.0\textwidth]{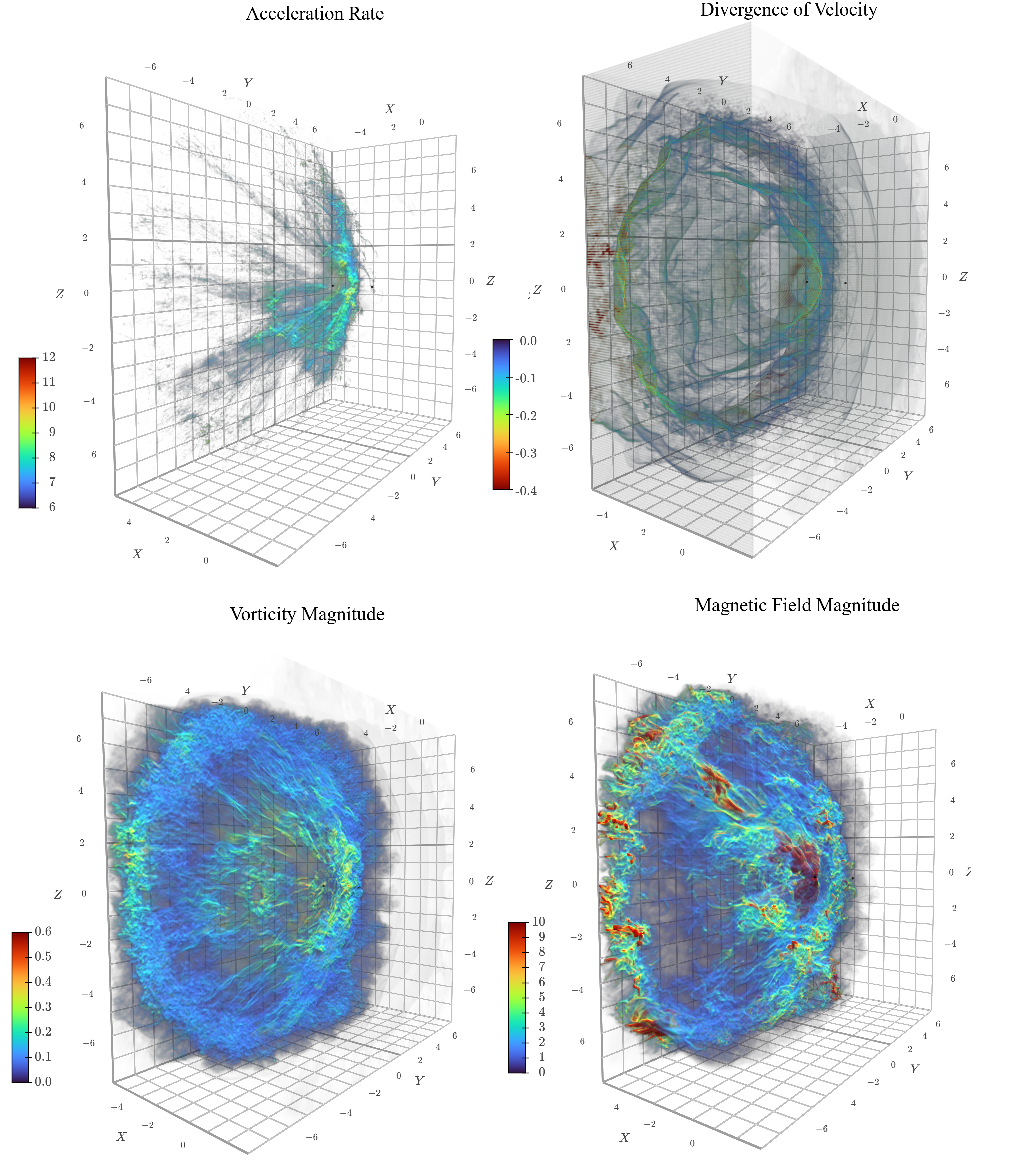}
	\caption{3-dimensional projection of: (top-left) the logarithm of the energy gain rate of the computed particles, in eV s$^{-1}$; (top-right) the divergence of velocity, $\bf{\nabla} \cdot \bf{u} $, in units of c/AU ; (bottom-left) vorticity of velocity, $\bf{\nabla} \times \bf{u} $, in units of c/AU; and (bottom-right) the magnetic field intensity, $|{\bf B}|$, in G; for the MHD model with $B_{\star} = 1$ G, $\Gamma_{\rm eff} = 5/3$ (adiabatic).}
	\label{mapsA}
\end{figure*}

\begin{figure*}
	\centering
	\includegraphics[width=1.0\textwidth]{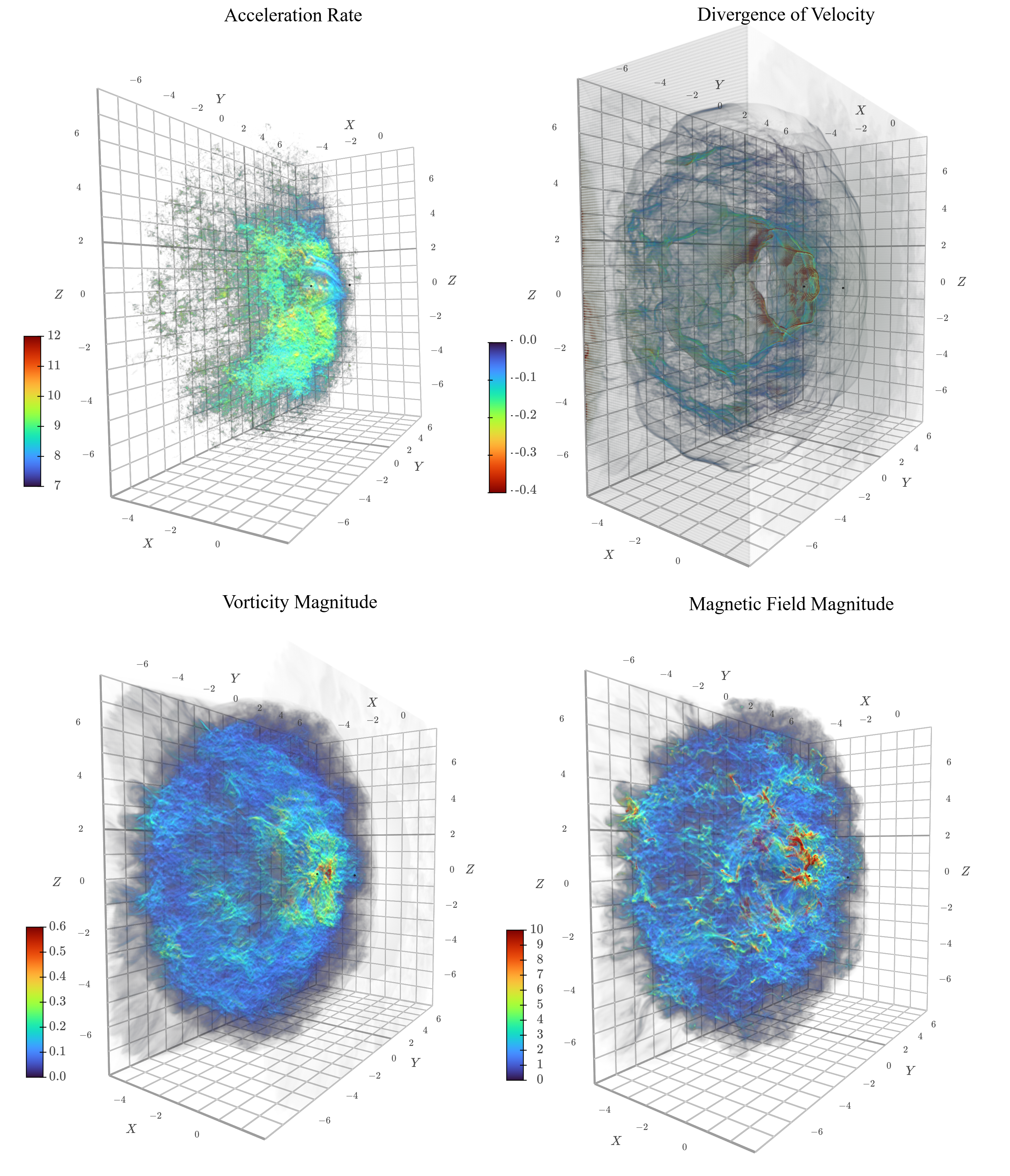}
	\caption{Same as Figure \ref{mapsA}, but for $\Gamma_{\rm eff} = 4/3$.}
	\label{mapsB}
\end{figure*}

\begin{figure*}
	\centering
	\includegraphics[width=1.0\textwidth]{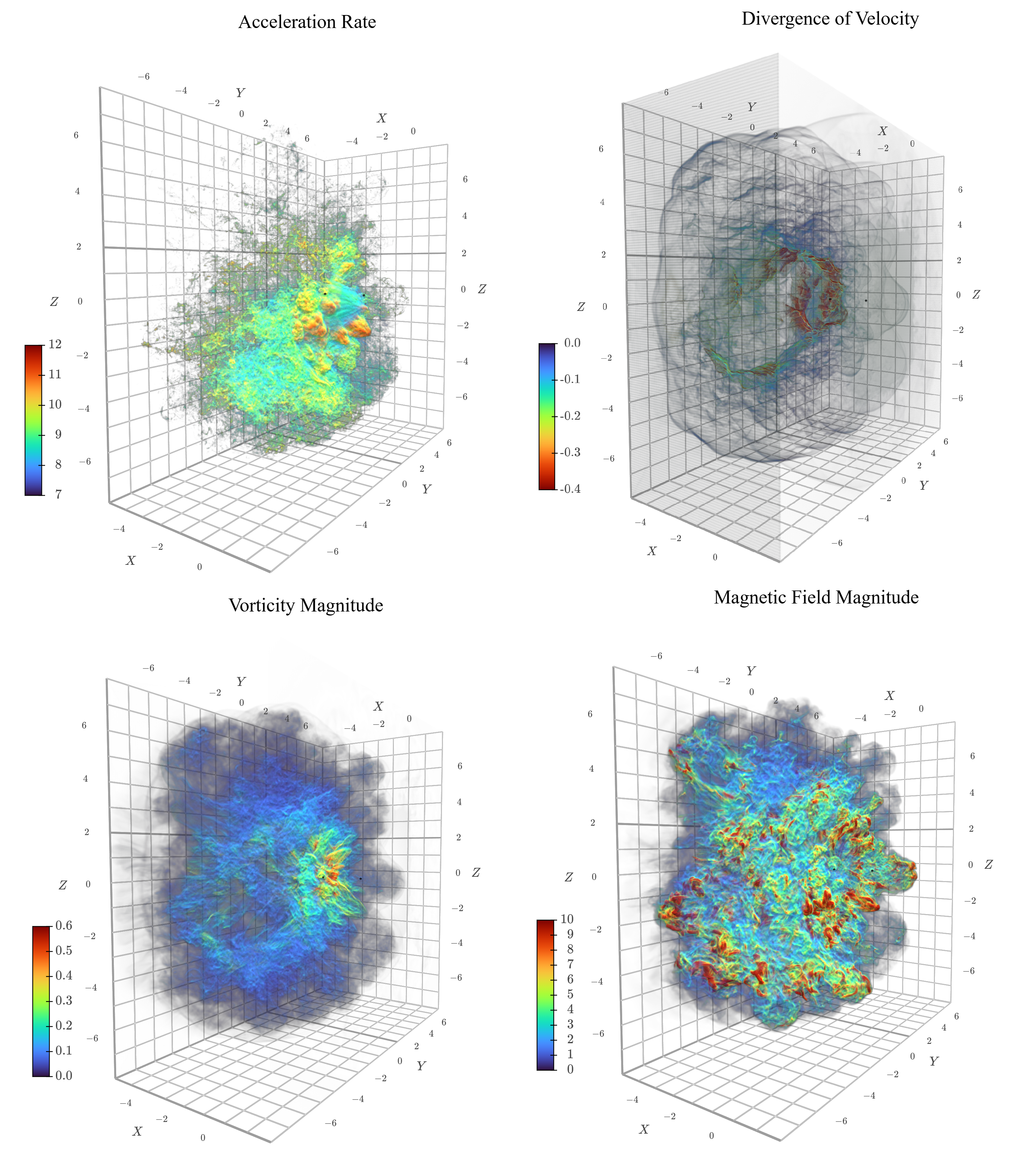}
	\caption{Same as Figure \ref{mapsA}, but for $\Gamma_{\rm eff} = 1.1$.}
	\label{mapsC}
\end{figure*}

{\it The shocks} -- A quantity that serves as a good indicator of shock locations is the divergence of the plasma velocity, $\nabla \cdot \mathbf{u}$. Maps of the negative values, which correspond to compression regions, are shown in the top-right panels of Figures~\ref{mapsA}, \ref{mapsB}, and \ref{mapsC} for $B_{\star} = 1$~G. The models represent distinct plasma compressibility regimes, with $\Gamma_{\rm eff} = 5/3$ (adiabatic), $4/3$, and $1.1$ (strong cooling), respectively.

The maps of negative values, corresponding to the compression regions, are shown in the right-top panels of Figures~\ref{mapsA},~\ref{mapsB} and ~\ref{mapsC}, for $B_{\star} = 1$ G. The models correspond to distinct plasma compressibility regimes, with $\Gamma_{\rm eff} = 5/3$ (adiabatic), $4/3$ and $1.1$ (strong cooling), respectively.

For $\Gamma_{\rm eff} = 5/3$, two shocks are clearly identifiable: one corresponding to the primary wind colliding with the downstream region, and the other corresponding to the secondary stellar wind colliding with the downstream region. These two shock layers enclose the dense shocked material where particle acceleration is expected to occur. The stronger shock (red region in the map) is the inner surface, associated with the shock of the secondary wind. Its conical shape, curved toward the secondary star, indicates that the momentum of the primary wind exceeds that of the secondary. However, since the secondary wind speed is higher than that of the primary, the shock is stronger for the secondary wind.

There is no substantial difference in the divergence-of-velocity distributions among the models. The two stars in the system are indicated by black dots in the figures.

{\it Turbulence} --  In the shocked gas, compression and shear in the velocity streams give rise to fluid instabilities that grow and non-linearly evolve into turbulent motions \citep[see][]{falceta2014b}. To trace the level of turbulence in the shocks, we use the vorticity of the plasma velocity field, $\boldsymbol\omega = \nabla \times \mathbf{u}$. The vorticity maps are shown in the bottom-left panels of Figures~\ref{mapsA}, \ref{mapsB}, and \ref{mapsC} for $B_{\star} = 1$ G.

It is noteworthy that the region where turbulence develops most strongly does not coincide with the location of the shocks (i.e., where $|\nabla \cdot \mathbf{u}|$ is maximum), but rather lies between the shock surfaces—in the downstream region of both shocks. The shocked gas from both winds is subject to the growth and evolution of instabilities that drive turbulence.

Another important aspect of the vorticity distributions is their dependence on $\Gamma_{\rm eff}$. The adiabatic model ($\Gamma_{\rm eff} = 5/3$) exhibits the weakest turbulent motions, corresponding to a smoother downstream flow. Models with stronger post-shock energy losses present higher compressibility and more intense turbulence. This effect, discussed previously \citep[see][]{falceta2012}, is linked to the enhanced growth of instabilities and the non-linear evolution of the downstream velocity field when the shocked gas undergoes rapid cooling.

{\it Magnetic Field} -- In terms of their level of magnetization, the shocks are responsible not only for the plasma compression (which makes the magnetic field increase with the density as $B \propto \rho$ in the downstream, but also the {\it pile-up effect} occurs \citep{falceta2012,fg2015,kowal.2021}. This effect is the consequence of the freezing of field lines, brought with the winds, on the contact discontinuity surface as the shocked plasma turns and flows along it. 

Typically, for a non turbulent downstream, the level of magnetization reaches its maximum at the contact discontinuity. In super-Alfvénic turbulent scenarios however the magnetic field is dragged by the shocked gas motions, filling the whole broader shock region. This is clearly visible in the bottom-right panels of Figures ~\ref{mapsA},~\ref{mapsB} and ~\ref{mapsC}, for $B_{\star} = 1$ G. As turbulence operates against the {\rm pile-up effect}, given that it depends on the smooth layout of field lines onto the contact discontinuity surface, redistributing the magnetic energy as turbulent Alfvén waves instead, the level of magnetization is maximum for the model with $\Gamma_{\rm eff} = 5/3$. However, the magnetic field is strong at a narrower region - the contact discontinuity surface -, mostly at the apex of the shock cone (as shown in Fig.~\ref{mapsA} (bottom-right). For the models where cooling occurs, the turbulent downstream is responsible for carrying the field lines along with the turbulent flow. The consequence is the reduction of the magnetization level at the apex of the shock, but the increase of the field intensity at the broad shocked region. As shown in Fig.~\ref{mapsC} (bottom-right), the magnetic field intensity is broadly distributed over the whole shocked region, and, being passive to the turbulent motions, its geometry is highly complex.

\subsection{Particle acceleration}

As described in Section \ref{sec:particles}, we integrate the trajectory of $10^5$ thermal particles, randomly positioned at the last snapshot of each of our MHD simulations, varying their mass-to-charge ratio in order to simulate different ion species interacting with the plasma and magnetic field of the CWB. At each position of the computational domain where a particle has passed through, its acceleration and energy are computed. In order to understand the physics of the acceleration of high energy particles in CWBs, one must correlate the acceleration rate to the physical properties of the plasma and the B-field, locally.  The logarithm of the energy gain rates are shown in the top-left panels of figures ~\ref{mapsA},~\ref{mapsB} and ~\ref{mapsC}.

The maximum energy and probability distributions of the different species do not vary much, as already demonstrated in \cite{kowal.2021}. For this reason, we will focus our analysis on H$^+$ nuclei only, which can be generalized to the other species.  

In the case with $\Gamma_{\rm eff} = 5/3$ (Figure \ref{mapsC}), regions where particles accelerate are located at the wind-wind shock region, as expected, and visually coincide with zones of strong vorticity, compression (i.e., negative velocity divergence), and enhanced magnetic field. This spatial overlap suggests the action of acceleration mechanisms that depend on the magnetic field topology, such as magnetic reconnection and the mirror effect, turbulent diffusion, and possible first order processes. For the different $\Gamma_{\rm eff}$´s, this scenario is modified. 

For $\gamma = 1.33$ (Figure \ref{mapsB}), the turbulent structures remain complex but are more spread. Regions of enhanced magnetic field coincide remarkably well with zones of high acceleration rate, indicating that magnetic field amplification contributes directly to the efficiency of the acceleration processes. Similarly to vorticity. The velocity divergence shows a compression pattern concentrated in the inner shock (of the secondary star), remarkably detached of the high acceleration rate zone.

Focusing now in the most turbulent case among the three models, for $\Gamma_{\rm eff} = 1.1$ (Figure \ref{mapsA}), we can notice that regions where the acceleration rates are very high (red regions) mostly coincide with turbulent and highly magnetized regions, with little to none spatial correlation to the regions of strongest shocks. Since this model is also the one that presents the highest acceleration rates, it reveals that turbulent diffusion and complexity of magnetic field geometry are key elements in the acceleration of particles in CWBs, while shock-related first order mechanisms, such as DSA, seem to play a minor role.

\subsubsection{Particle Energy distributions}

\begin{figure*}
\centering
\includegraphics[width=0.45\textwidth]{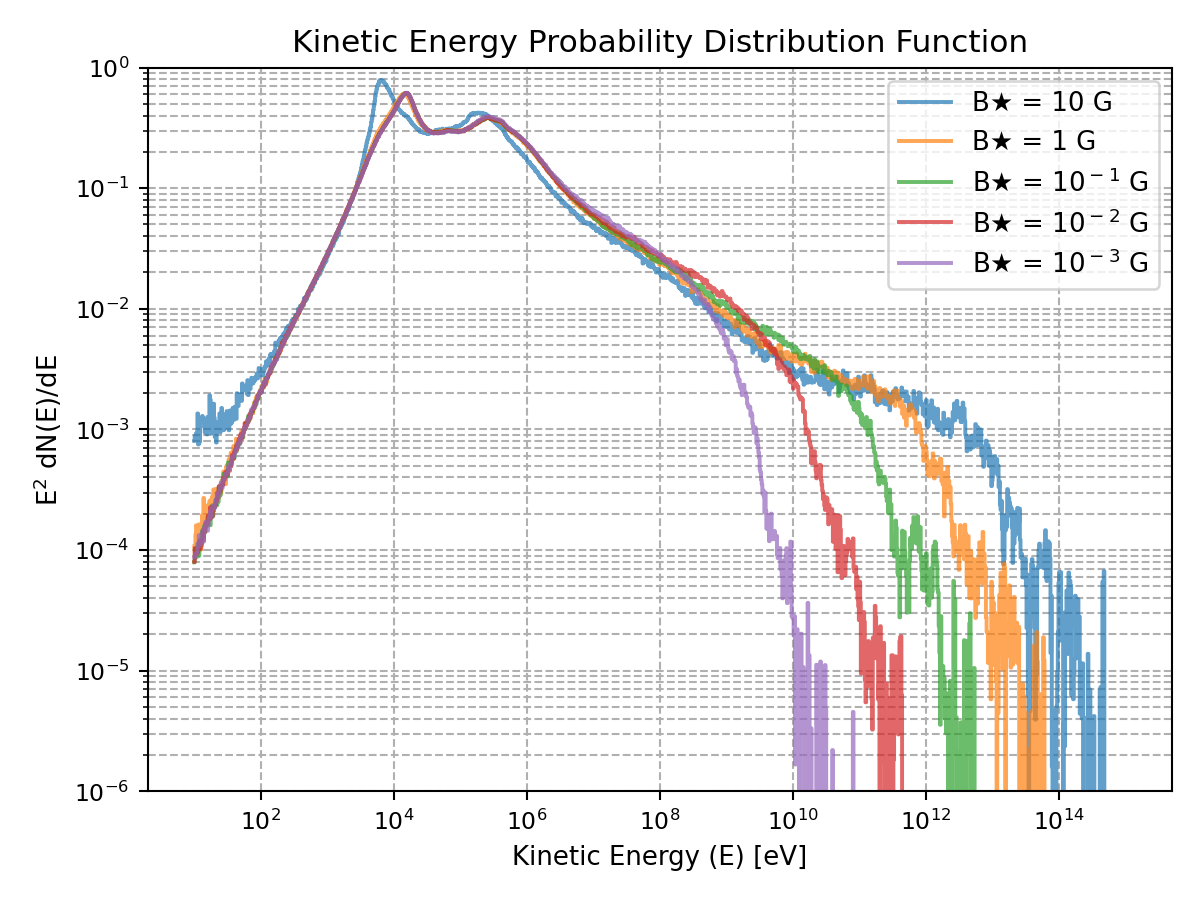}
\includegraphics[width=0.45\textwidth]{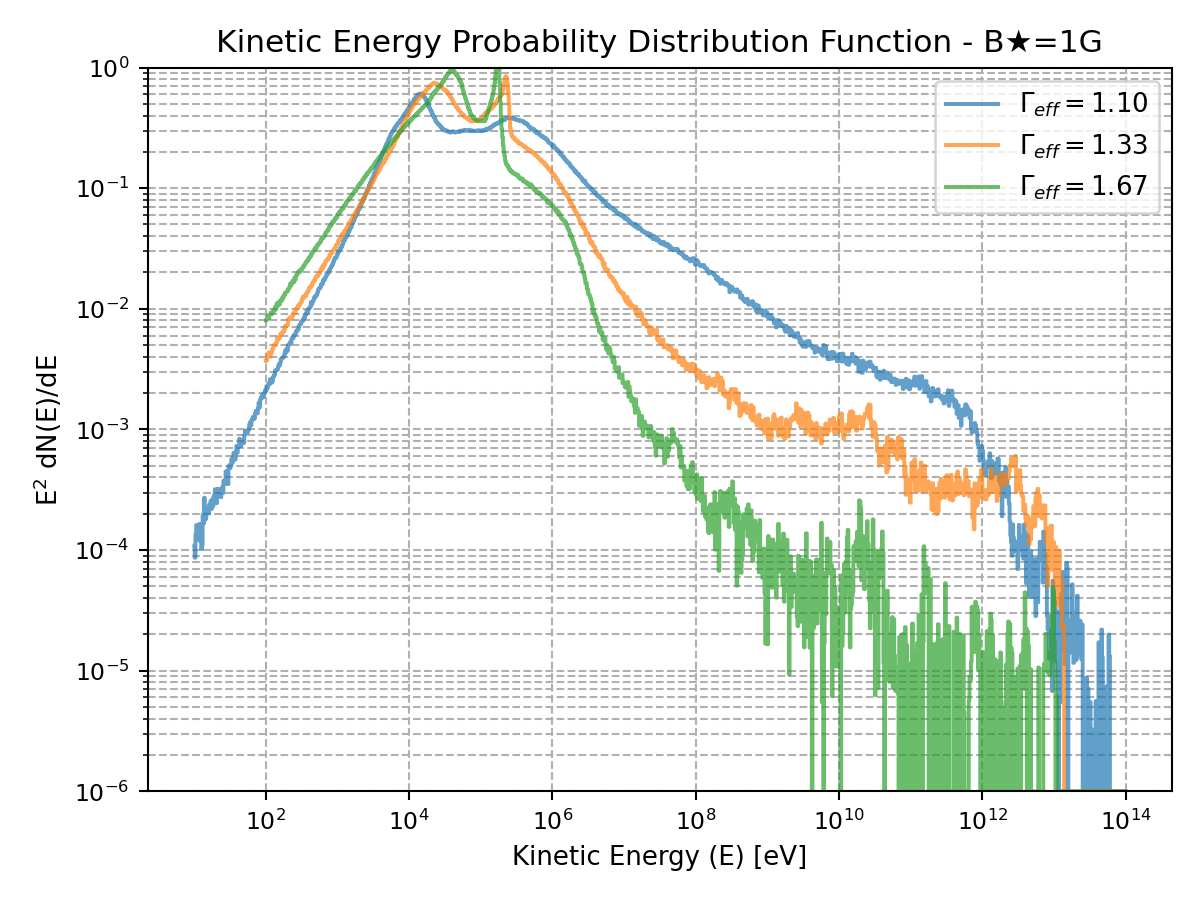}
\includegraphics[width=0.45\textwidth]{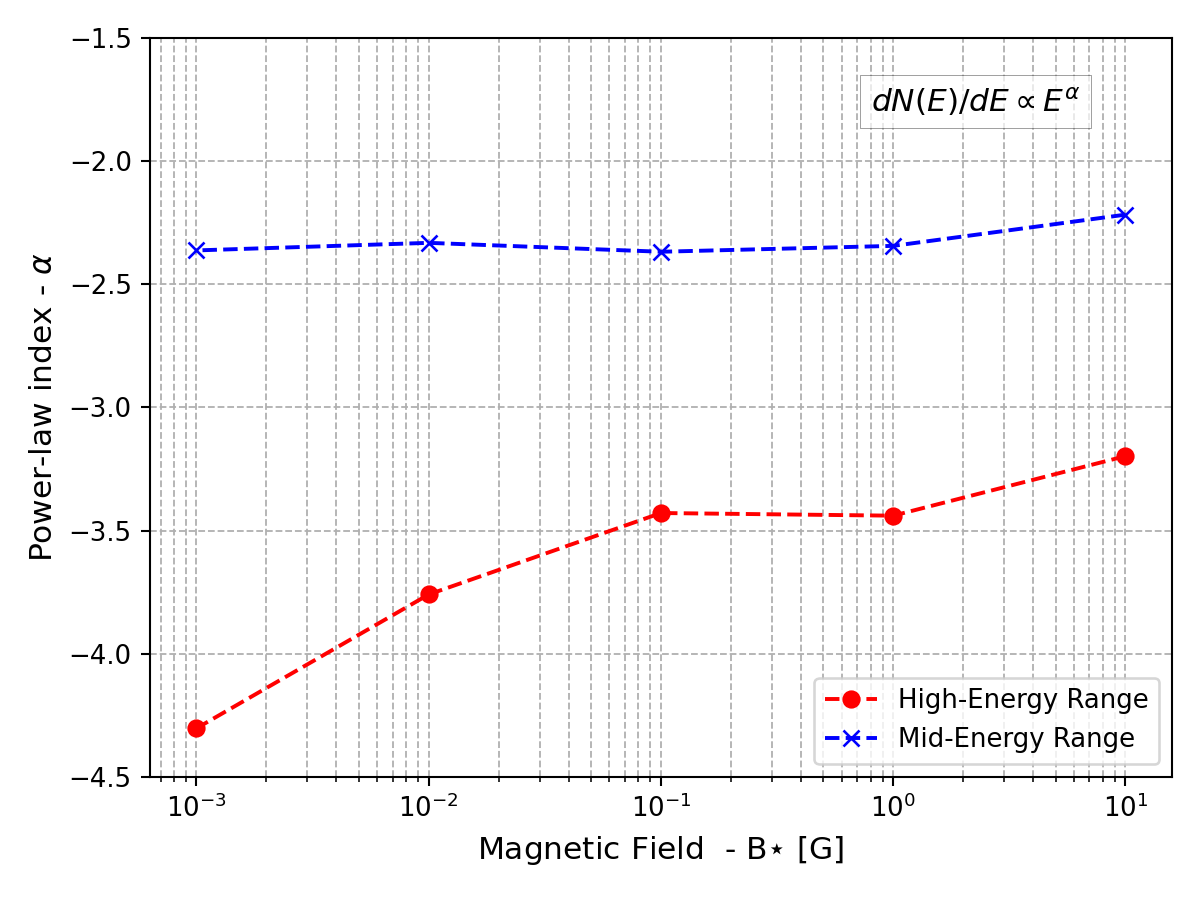}
\includegraphics[width=0.45\textwidth]{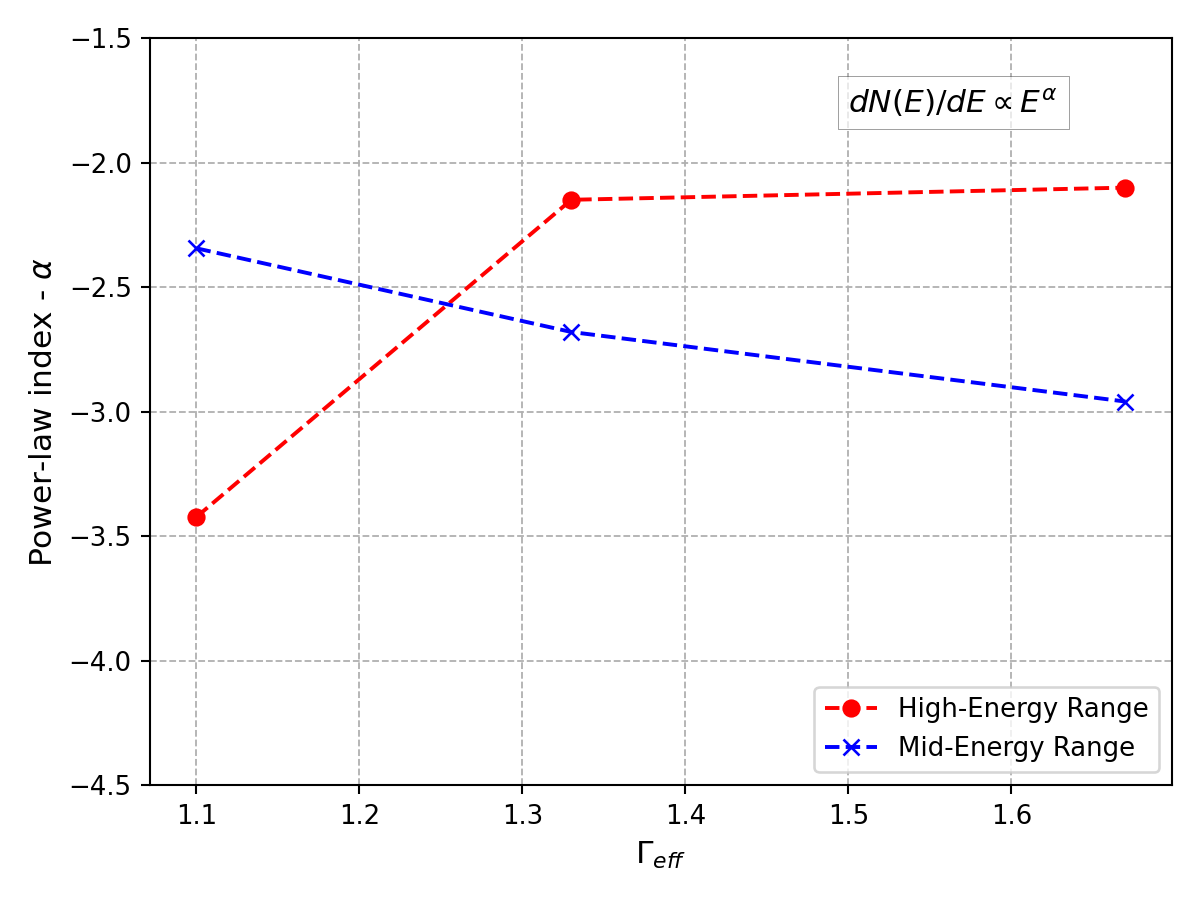}
\caption{Top-left: Particle energy distribution multiplied by $E^2$, for the models with distinct values of $B_{\star}$, and $\Gamma_{\rm eff} = 1.1$. For the mid and high-energy ranges (see text for details), separated by the high energy break, two power-laws of $dN(E)/dE \propto E^{\alpha}$ were fitted.  Bottom-left: The two values of $\alpha$ for each energy range ($\times$-symbol for the mid-energy range, and $\circ$-symbol for the high-energy range) are shown as a function of the magnetic field $B_{\star}$. Top-right: Particle energy distribution multiplied by $E^2$, for the models with distinct values of $\Gamma_{\rm eff}$, and $B_{\star} = 1$G. Bottom-right: The two values of $\alpha$ as a function of $\Gamma_{\rm eff}$ for $B_{\star} = 1$G.}
\label{PDF}
\end{figure*}

In Figure~\ref{PDF}, we present the particle energy distribution (multiplied by $E^2$), for the models with distinct values of $B_{\star}$ (top-left), and $\Gamma_{\rm eff}$ (top=right). In the top-left panel, all distributions obtained may be characterized as a composition of three segments, each with its characteristic slope: one at low-energy range ($E < 10^5$eV), and the mid and high-energy ranges separated by a break hereby denominated as the {\it knee}. The low-energy range of the particle energy  distribution ($E < 10^5$eV) is of little interest here as it is determined solely by the initial thermal distribution of particles, which originate at the hot plasma at the shocks. The mid and high-energy ranges of the distributions correspond to particles with energies altered by the interaction with the plasma dynamics and magnetic fields, i.e. those that suffered acceleration.

One important result obtained is that the maximum energy of particles grows as $B_{\star}$ increases. This result is in agreement to those found in \citet{kowal.2021}, who found a linear relationship between $B_{\star}$ and the maximum energy of particles, though for a different set of simulations, and with coarser resolution. 
Not only the cutoff energy is dependent on the magnetic field intensity, but the locus of the energy break at the higher energies is also dependent on $B_{\star}$. This means that the break in the energy distribution is not a consequence of poor statistics of the number of particles integrated, but carries a physical meaning: the trapping of particles within the shock - i.e. the Hillas confinement
criterion. The ``knee'' observed in the energy PDF possibly characterizes the transition of propagation regime of the particles. Above this energy, particles may escape depending on the location they are accelerated. Since the lengthscales of the turbulence and the magnetic field intensity of the fluctuations vary corresponding to a spectrum of values, the escape occurs in a range of energy, instead of as a sharp cutoff. This explain the steeper slope of the energy PDF at energies beyond the break. The magnetization level does not seem to be relevant for the shape of the particle energy distribution. On the other hand, the polytropic index $\Gamma_{\rm eff}$ shows strong effect in the energy PDF.

In the top-right panel of Fig.~\ref{PDF}, the particle energy distributions obtained for different $\Gamma_{\rm eff}$, for a fixed $B_{\star}=1$G, reveal a softer distribution in the mid-energy range as the shocks become more adiabatic. In the high energy range though, this trend is reversed and the distribution becomes harder as $\Gamma_{\rm eff}$ increases. This different trend does not modify the maximum energy achieved by particles, but result in a convergence of the PDFs at the highest energies. Again, this result is in agreement to the Hillas criterium, given that the typical magnetic field intensities within the shocks of the CWB depend mostly on $B_{\star}$, and only marginally on $\Gamma_{\rm eff}$. 

By separating the mid and high-energy ranges around the knee, we were able to perform linear log-log regression fitting of the PDF, in the form of two $dN(E)/dE \propto E^{\alpha}$, i.e. broken power-law. The values of $\alpha$ for each model, for both energy ranges around the knee, are shown in the bottom plots of Figure~\ref{PDF}. In the bottom-left panel, $\alpha$ parameter of the mid-energy range is similar ($\sim -2.4$ to $-2.2$) for all values of $B_{\star}$. For the high-energy range however the slope is related to level of magnetization, varying from $-4.4$ to $-3.2$ as $B_{\star}$ increases. With respect to the polytropic index, for a fixed $B_{\star} = 1$G, we find that as $\Gamma_{\rm eff}$ increases from 1.1 to 5/3 the slope $\alpha$ decreases from $\sim -2.3$ to $\sim -3.0$ in the mid-energy range, but increases from $\sim -3.4$ to $\sim -2.1$ in the high-energy regime. 

The absence of a clear dependence of $\alpha$ on the level of magnetization at the mid-energy range shows that the same mechanism operates to accelerate particles within this range. The position of the break, as mentioned before, determines the change in regime where confined particles start to diffuse out of acceleration zones, reducing its effectiveness, and resulting in steeper values of $\alpha$. 

Cooling effects at the shock region revealed to be extremely relevant for the acceleration of particles in the CWB scenario, as the polytropic index, $\Gamma_{\rm eff}$ determines the compressibility level - and therefore the clumpiness - of the shock region. Adiabatic shocks present less complexity of topology of the magnetic field, therefore less regions where particles may be trapped. This fact might be the explanation for the steeper $\alpha$ value and the lower number of particles accelerated, indicating a less efficient acceleration, compared to what is found in the models with lower values of $\Gamma_{\rm eff}$. The flattened high-energy range of the distribution, with slope $\alpha \sim -2$, indicates that a shock-mediated mechanism (such as DSA) is present and overcomes the turbulent magnetic process at these energies.

\subsection{Correlation between physical properties and particle acceleration}

In this Section we analyze the relationship between the particle acceleration rate and various plasma properties. The quantities considered are: the local magnetic field, current density, velocity, vorticity, and velocity divergence, all extracted directly from the three-dimensional snapshots of the simulation. The aim of this analysis is to identify the regions most favorable to efficient energy gain by initially thermal particles, a key mechanism in the production of high-energy cosmic rays.

We employed two-dimensional probability distribution functions (2D-PDF) to correlate the energy gain rate $(\frac{dE}{dt})$ with the  physical properties of the plasma and environment B-field, allowing for a global statistical assessment of the system's behavior. This analysis was performed for the whole set of simulations, i.e. varying both the boundary magnetic field intensity $(B^{\star})$ and the effective polytropic index $(\Gamma_{\rm eff})$, representing different plasma compressibility regimes. 
We then statistically obtain, for each 2D-PDF, the Pearson correlation coefficient as well as the linear fit regression function. To ensure that only regions with significant acceleration were included in these calculations, we applied a cutoff threshold considering only the regions where the energy gain rate exceeds 100 eV/s.

\begin{figure*}   
  \centering
  \includegraphics[width=\textwidth]{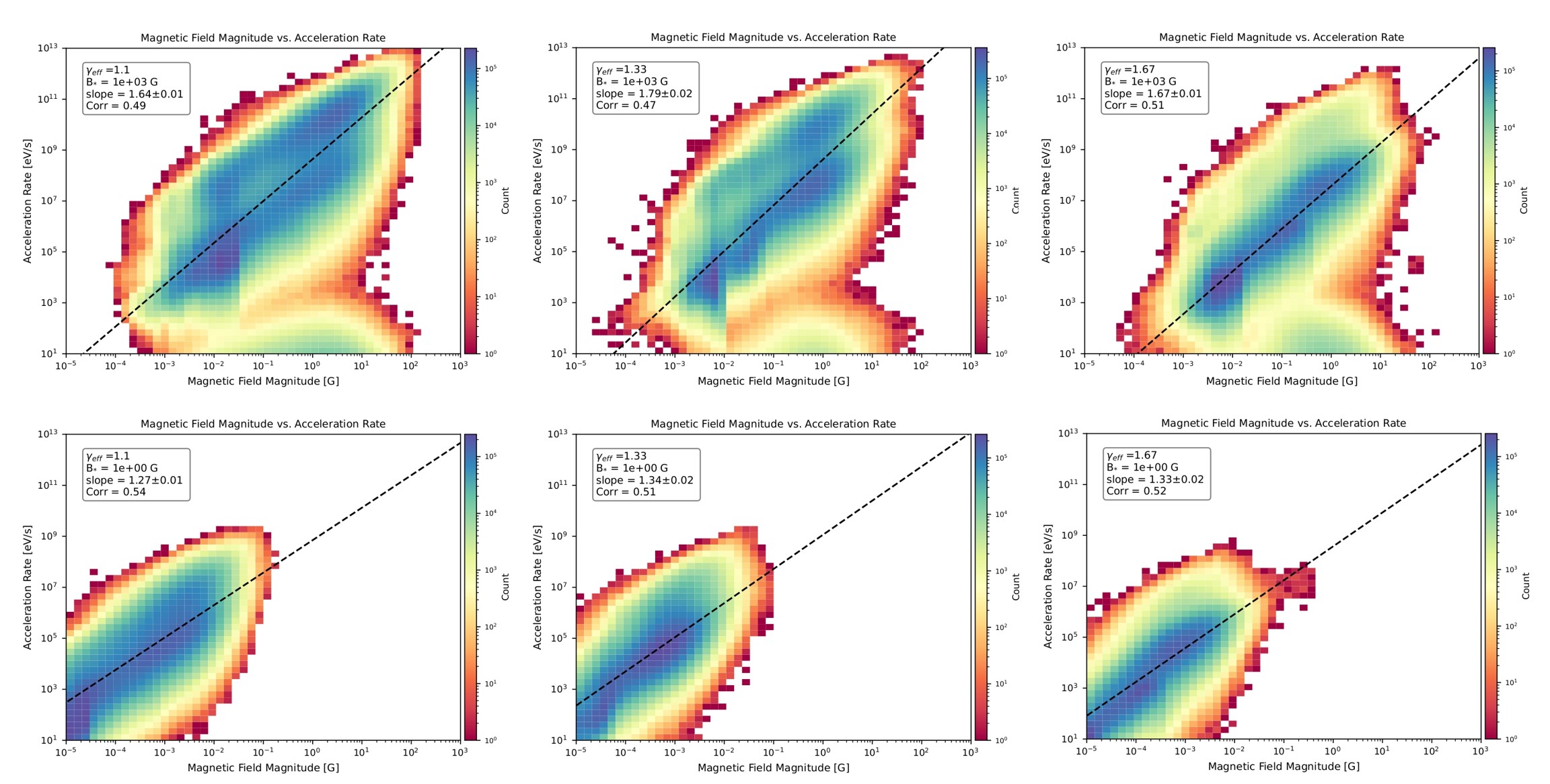}
  \caption{Correlations between magnetic field strength and particle acceleration rate for different polytropic indices ($\Gamma_{\rm eff}$) and magnetic field intensities (B$_{\star}$). Each panel shows a logarithmic 2D-PDF, representing the density of points in the $\lvert \vec{B} \rvert$ × $\frac{dE}{dt}$ phase space. The horizontal axis indicates the magnetic field intensity of the regions where acceleration occur, and the vertical axis shows the energy gain rate (in eV/s). The black dashed line represents the statistical linear regression for the data, with the fitted correlation index indicated in the top-left corner of each panel. Pearson correlation values are also displayed. Comparisons were made for the different models, with two values of B$_{\star}= 1$ and $10^3$ G, and different polytropic indices ($\Gamma_{\rm eff}$ = 1.10, 1.33, and 1.67). }
  \label{magxacc}
\end{figure*}

In Figure~\ref{magxacc} we present the 2D-PDF correlating the particle energy gain rate ($dE/dt$) with the magnetic field intensity ($|\mathbf{B}|$), of the regions where particles are accelerated. The different panels correspond to different $B_{\star}$ and polytropic index. The upper panels correspond to $B_{\star}  = 10^3$G, while bottom panels to $B_{\star}  = 1$G. From left to right, $\Gamma_{\rm eff} = 1.1, 1.33$ and 1.67, respectively. We focused the estimation of the correlations only for regions where the energy gain rate exceeds 100 eV/s. A quite strong positive correlation is observed between the variables, with Pearson correlation coefficients ranging from 0.47 to 0.54. 

The correlation coefficients of the 2D-PDFs, obtained via logarithmic-scale linear regression, range from 1.27 to 1.79, revealing a clear pattern: in simulations with stronger initial magnetic fields ($B_\star = 10^3$ G), the slopes are higher (up to 1.79), while in cases with weaker fields ($B_\star = 1$ G), the slopes decrease (between 1.27 and 1.34). This superlinear behavior indicates high correlation between $dE/dt$ and $|\mathbf{B}|$. Changes in the polytropic index $\Gamma_{\rm eff}$ however do not lead to significant variations in this correlation. An explanation would be that the degree of plasma compressibility — associated with the orbital phase in the binary system — affects the global shock structure and geometry, but has a secondary effect on the magnetic field - acceleration relationship, given that it is a local (small scale) process.

From a theoretical perspective, the central role of the magnetic field is supported by the generalized Ohm’s law in MHD: ${\bf E} + {\bf u} \times {\bf B} = \eta {\bf J}$. Therefore, the effects of the magnetic field in accelerating the particles may occur from a combined mechanism with the fluid motion, or thought the current density.

\begin{figure*}   
  \centering
  \includegraphics[width=\textwidth]{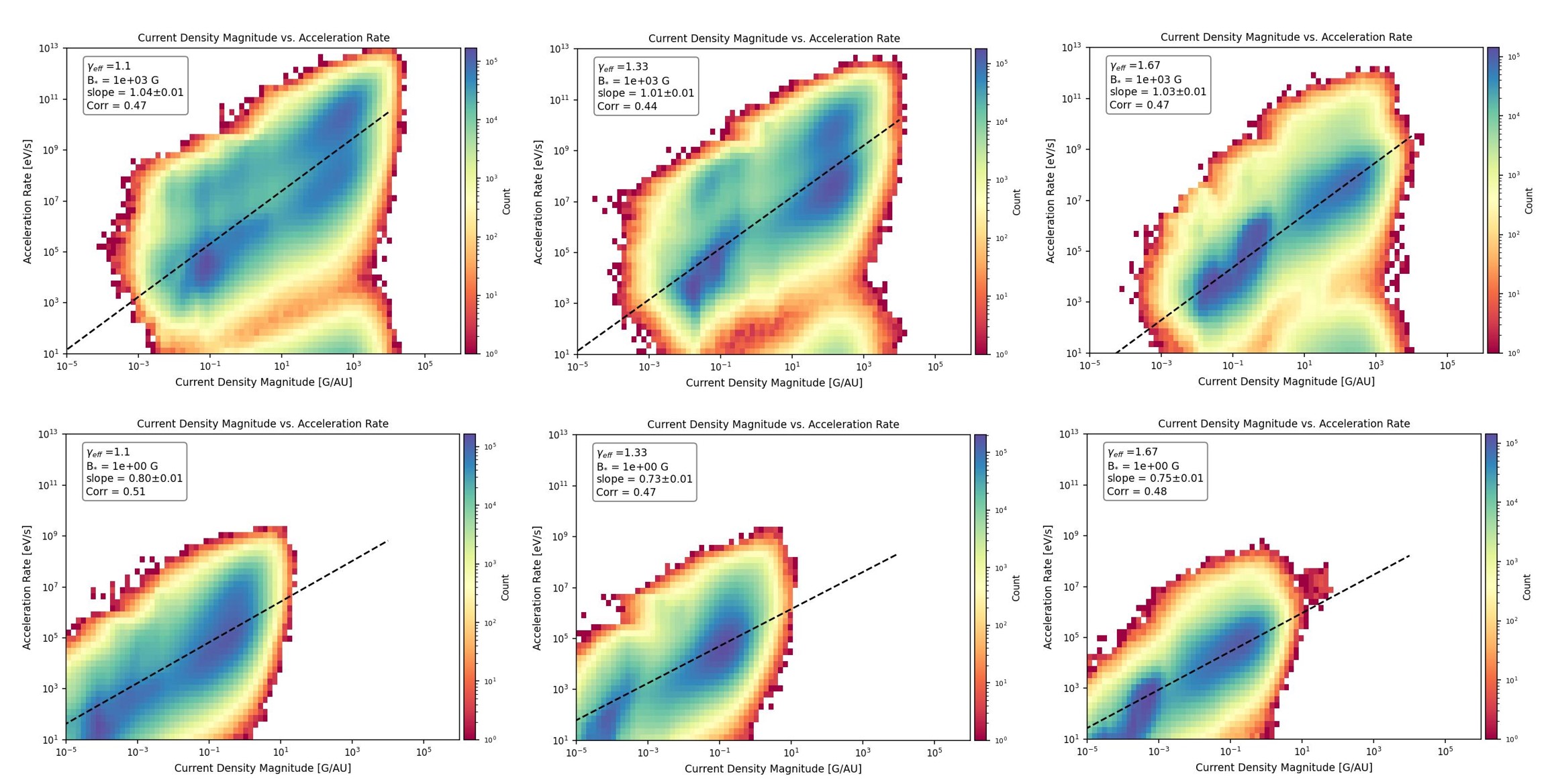}
  \caption{Same as shown in Figure \ref{magxacc} but for correlations between current density and particle acceleration rate. 
  }
  \label{cdxacc}
\end{figure*}

Firstly, let us analyze the correlation between the acceleration with the current density. Figure~\ref{cdxacc} presents the 2D-PDF of $dE/dt$ and the current density magnitude ($|\mathbf{J}|$). The different panels correspond to different $B_{\star}$ and polytropic index. The upper panels correspond to $B_{\star}  = 10^3$G, while bottom panels to $B_{\star}  = 1$G. From left to rigth, $\Gamma_{\rm eff} = 1.1, 1.33$ and 1.67, respectively. As in the previous analyses, only the cells with acceleration rates above 100 eV/s were considered, focusing on the most active regions of the simulation. Pearson correlation coefficients range from 0.44 to 0.51, indicating a moderate association between these variables, with a statistical behavior similar to that observed for the magnetic field.

However, compared to the correlation found for $|\mathbf{B}|$, the coefficients obtained are lower, ranging from 0.73 to 1.04. This can be explained from the fact that $\eta$ is virtually negligible in most astrophysical environments, and $\eta {\bf J}$ is much smaller compared to its dynamic counterpart in the Ohm´s law, $-{\bf v} \times {\bf B}$.
Although current density shows a statistically significant correlation with $dE/dt$, its direct contribution to the acceleration process is limited within the adopted model, being such correlation indirectly determined by the magnetic field intensity itself (given that for virtually ideal MHD, $J \propto {\bf \nabla \times B}$). The distribution pattern is similar to that observed in $|\mathbf{B}|$, reinforcing the idea that regions of high current density tend to spatially coincide with regions of strong magnetic field — but are not necessarily the direct cause of acceleration. In this context, the current density acts as an indirect tracer of the complexity of the magnetic field, rather than a direct controller of acceleration rates.

Since the resistive term $\eta$ in the simulations (and in nature) is extremely small, the dominant contribution arises from the magnetic term rather than from resistive driven electric fields. Therefore, particle acceleration is directly influenced by fluctuations in the magnetic field and plasma flow, favored in turbulent regions with small scale spatial variations in $|\mathbf{B}|$. 
Thus, while the magnetic field alone plays a crucial role in regulating acceleration efficiency, the results point towards a combined action  — particularly in regions of strong turbulence, as will be studied below. 

\begin{figure*}   
  \centering
  \includegraphics[width=\textwidth]{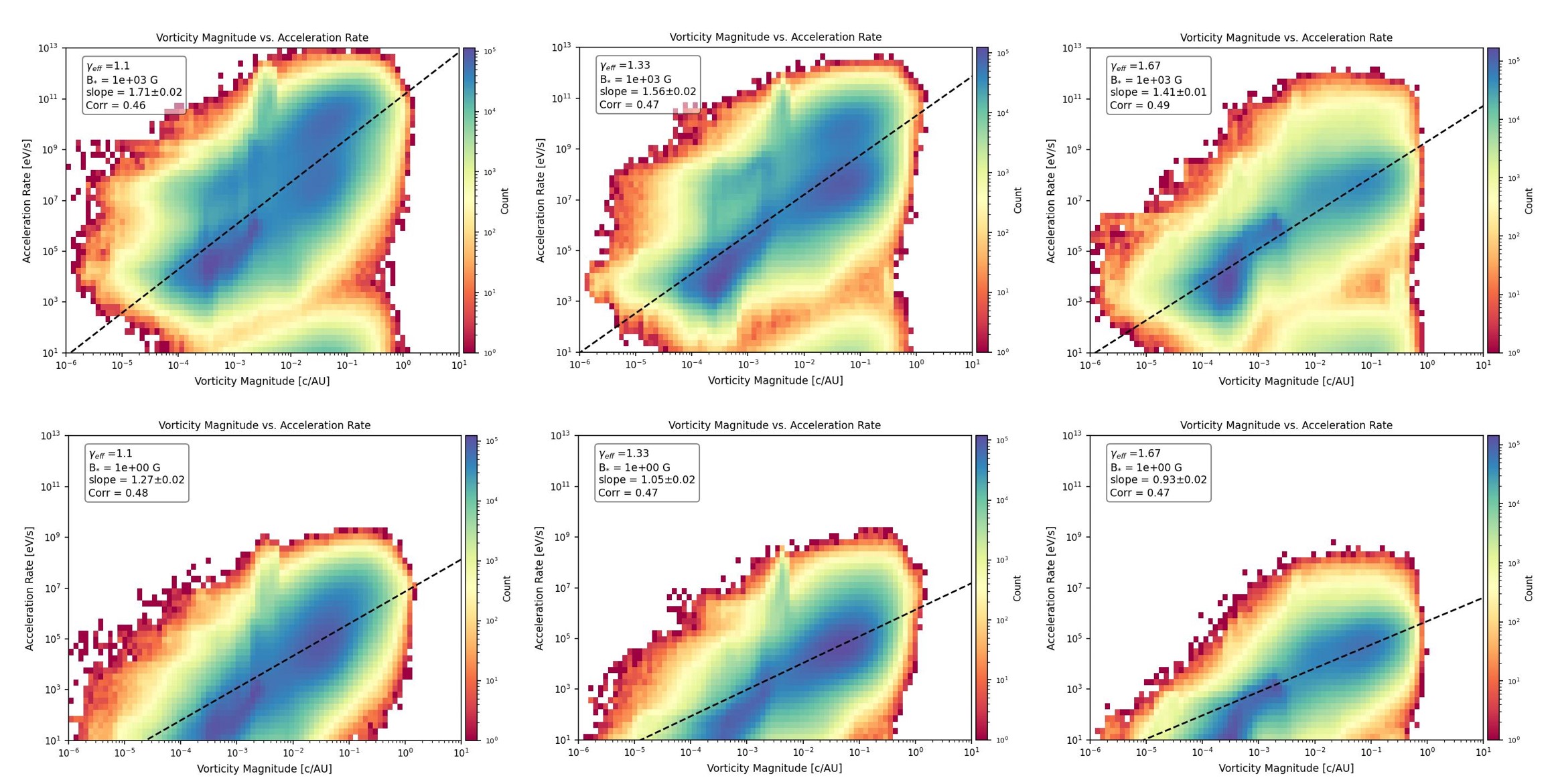}
  \caption{Same as shown in Figure \ref{magxacc} but for correlations between vorticity of the velocity field and particle acceleration rate.}
  \label{vortxacc}
\end{figure*}

Figure~\ref{vortxacc} presents two-dimensional histograms relating the particle energy gain rate ($dE/dt$) to the vorticity magnitude ($|{\omega}|$). The different panels correspond to different $B_{\star}$ and polytropic index. The upper panels correspond to $B_{\star}  = 10^3$G, while bottom panels to $B_{\star}  = 1$G. From left to rigth, $\Gamma_{\rm eff} = 1.1, 1.33$ and 1.67, respectively. As in the previous analyses, only the cells with energy gain rates above 100 eV/s were considered. Pearson correlation coefficients range from 0.46 to 0.49, indicating a moderate and consistent correlation between vorticity and acceleration rate — slightly lower than that observed for the magnetic field and similar to the current density.
The regression coefficients are in a range from 0.93 to 1.71, with higher values obtained for stronger magnetic fields ($B_\star = 10^3$ G) and lower $\gamma_{\rm eff}$ (i.e., high compressibility). This suggests that the particles acceleration rates are higher in more turbulent and magnetized regions. 

Physically, vorticity measures the local rotation of the fluid and is closely associated with turbulence. Highly vortical flows promote particle trapping, increasing their residence time within acceleration zones and enhancing stochastic mechanisms. Super-Alfvénic turbulence is also responsible for distorting the magnetic field, boosting the perpendicular velocity component of the particle relative to the field, thereby strengthening the effect of the Lorentz force.

The fact that the distributions exhibit significant scatter — especially at intermediate vorticity values — reinforces the hypothesis that multiple and interactive acceleration mechanisms are at play. Vorticity traces favorable conditions for inductive and stochastic processes to occur. In this sense, its contribution to acceleration occurs not only by ${\bf v} \times {\bf B}$ term, but also through its topological influence on the geometry of the flow and the magnetic field, such as the {\it betatron} effect and converging magnetic mirrors.

\begin{figure*}   
  \centering
  \includegraphics[width=\textwidth]{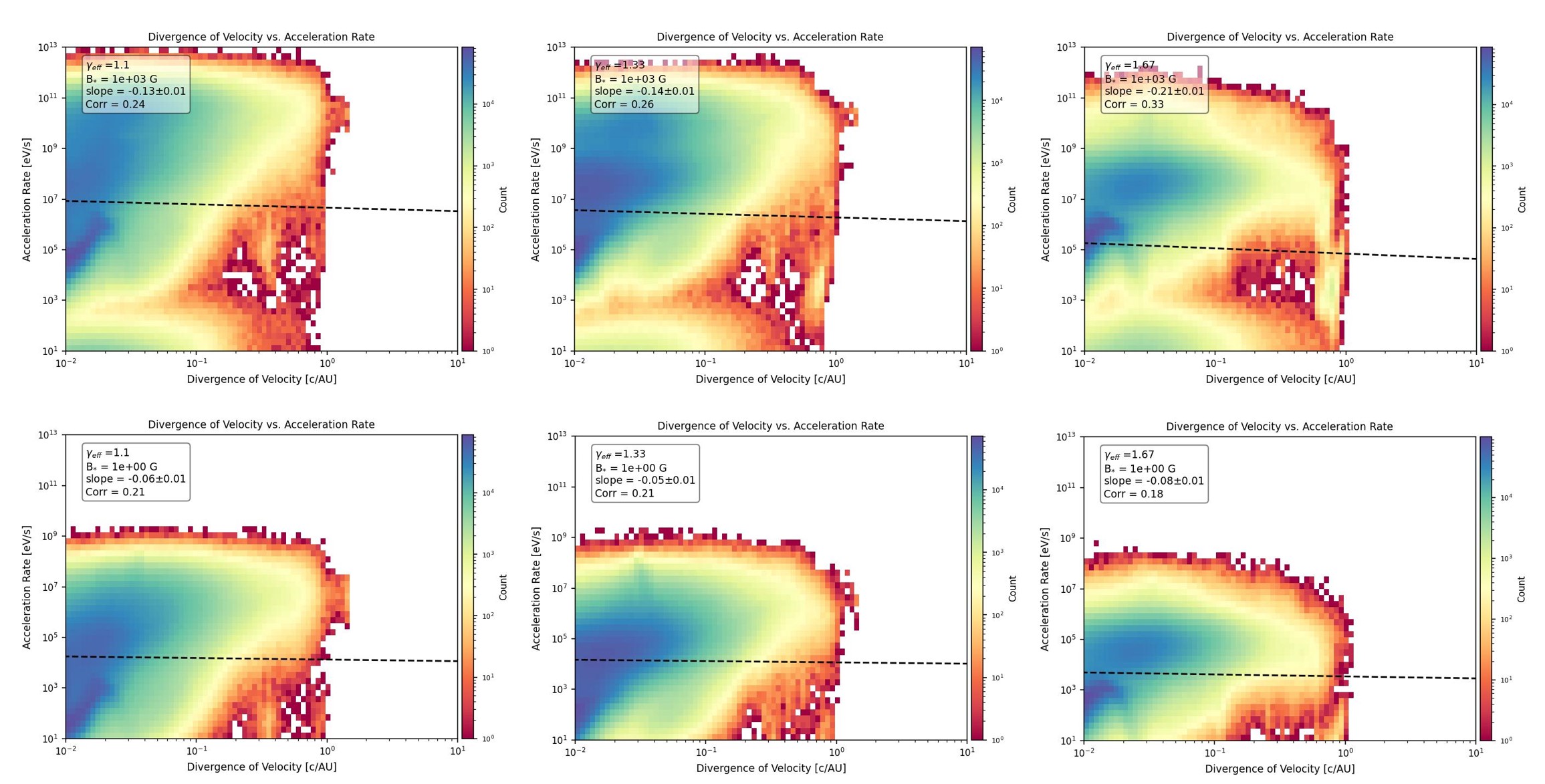}
  \caption{Same as shown in Figure \ref{magxacc} but for correlations between the divergence of velocity and particle energy gain rate.}
  \label{divxacc}
\end{figure*}

Now, we study the direct role of shock-induced first-order acceleration mechanism in the particle energy gain rates obtained from our simulations. Figure~\ref{divxacc} presents two-dimensional histograms relating the particle energy gain rate ($dE/dt$) to the divergence of the fluid velocity ($| \nabla \cdot {\bf v} |$). The different panels correspond to different $B_{\star}$ and polytropic index. The upper panels correspond to $B_{\star}  = 10^3$G, while bottom panels to $B_{\star}  = 1$G. From left to right, $\Gamma_{\rm eff} = 1.1, 1.33$ and 1.67, respectively. Again, we concentrate our analysis to regions with energy gain rates above 100 eV/s.

The Pearson correlation coefficients obtained here are the lowest among all the physical quantities analyzed, ranging from 0.18 to 0.33. Moreover, all slope values are negative, varying from -0.05 to -0.21. This pattern indicates a weak to no-correlation between the velocity field divergence and the particle acceleration rate.

\begin{table*}
  \centering
  \caption{Pearson correlation coefficients ($r$) between the proton acceleration rate and the fluid quantities obtained from the magnetohydrodynamic simulations. Three polytropic indices were considered ($\gamma$ = 1.10, 1.33, 1.67), along with different guide field intensities ($B^{} = 1$, $10^{1}$, $10^{2}$, $10^{3}$) in Gauss.}
  \small
  \setlength\tabcolsep{4pt}
  \renewcommand{\arraystretch}{1.15}
  \begin{tabular}{
      l
      *{4}{S[table-format = +1.2]} 
      *{2}{S[table-format = +1.2]} 
      *{4}{S[table-format = +1.2]} 
  }
    \toprule
    \multirow{2}{*}{} &
    \multicolumn{4}{c}{$\gamma = 1.10$} &
    \multicolumn{2}{c}{$\gamma = 1.33$} &
    \multicolumn{4}{c}{$\gamma = 1.67$} \\
    \cmidrule(lr){2-5}\cmidrule(lr){6-7}\cmidrule(lr){8-11}
      & {$B^{*}=10^{3}$}
      & {$B^{*}=10^{2}$}
      & {$B^{*}=10^{1}$}
      & {$B^{*}=1$}
      & {$B^{*}=10^{3}$}
      & {$B^{*}=1$}
      & {$B^{*}=10^{3}$}
      & {$B^{*}=10^{2}$}
      & {$B^{*}=10^{1}$}
      & {$B^{*}=1$} \\
    \midrule
    Magnetic Field $|\mathbf{B}|$ & +0.49 & +0.52 & +0.55 & +0.54 & +0.47 & +0.51 & +0.51 & +0.52 & +0.49 & +0.52 \\
    Vorticity $|\nabla\times\mathbf{v}|$ & +0.46 & +0.40 & +0.42 & +0.48 & +0.47 & +0.47 & +0.49 & +0.50 & +0.51 & +0.47 \\
    Current Density $|\mathbf{J}|$ & +0.47 & +0.47 & +0.50 & +0.51 & +0.44 & +0.47 & +0.47 & +0.49 & +0.48 & +0.48 \\
   Divergence of velocity $\nabla\!\cdot\!\mathbf{v}$
         & +0.24 & +0.17 & +0.18 & +0.21 & +0.21 & +0.26 & +0.33 & +0.32 & +0.27 & +0.18 \\
    \bottomrule
  \end{tabular}
  \label{tab:pearson}
\end{table*}

\begin{table*}
  \centering
  \caption{Regression coefficients and their respective uncertainties obtained from log–log fits for correlation between the acceleration rate and each fluid quantity, for the same simulation sets with three polytropic indices ($\gamma$) and four magnetic field strengths ($B^{}$), given in Gauss.}
  \small
  \setlength\tabcolsep{3pt}
  \renewcommand{\arraystretch}{1.15}

\begin{tabular}{
    l          
    S @{$\,\pm\,$} S
    S @{$\,\pm\,$} S
    S @{$\,\pm\,$} S
    S @{$\,\pm\,$} S
    S @{$\,\pm\,$} S
    S @{$\,\pm\,$} S
    S @{$\,\pm\,$} S
    S @{$\,\pm\,$} S
    S @{$\,\pm\,$} S
    S @{$\,\pm\,$} S
}
  \toprule
  \multirow{2}{*}{} &
    \multicolumn{8}{c}{$\gamma = 1.10$} &
    \multicolumn{4}{c}{$\gamma = 1.33$} &
    \multicolumn{8}{c}{$\gamma = 1.67$} \\
  \cmidrule(lr){2-9}\cmidrule(lr){10-13}\cmidrule(lr){14-21}
&   \multicolumn{2}{c}{$B^{*}=10^3$} &
    \multicolumn{2}{c}{$B^{*}=10^{2}$} &
    \multicolumn{2}{c}{$B^{*}=10^{1}$} &
    \multicolumn{2}{c}{$B^{*}=1$} &
    \multicolumn{2}{c}{$B^{*}=10^3$} &
    \multicolumn{2}{c}{$B^{*}=1$} &
    \multicolumn{2}{c}{$B^{*}=10^3$} &
    \multicolumn{2}{c}{$B^{*}=10^{2}$} &
    \multicolumn{2}{c}{$B^{*}=10^{1}$} &
    \multicolumn{2}{c}{$B^{*}=1$} \\[2pt]
  \midrule
  
    Magnetic Field $|\mathbf{B}|$ 
         & 1.64 & 0.01 & 1.28 & 0.01 & 1.42 & 0.01 & 1.27 & 0.01
         & 1.79 & 0.02 & 1.34 & 0.02
         & 1.67 & 0.01 & 1.63 & 0.01 & 1.48 & 0.01 & 1.33 & 0.02 \\
    Vorticity $|\nabla\times\mathbf{v}|$ & 1.70 & 0.02 & 1.08 & 0.01 & 1.13 & 0.01 & 1.27 & 0.02
         & 1.53 & 0.02 & 1.05 & 0.02
         & 1.41 & 0.01 & 1.35 & 0.01 & 1.16 & 0.01 & 0.93 & 0.02 \\
    Current Density $|\mathbf{J}|$ & 1.04 & 0.01 & 0.77 & 0.01 & 0.83 & 0.01 & 0.80 & 0.01
         & 1.01 & 0.01 & 0.73 & 0.01
         & 1.03 & 0.01 & 1.00 & 0.01 & 0.89 & 0.01 & 0.75 & 0.01 \\
   Divergence of Velocity $\nabla\!\cdot\!\mathbf{v}$
         & -0.13 & 0.01   & -0.06 & 0.01   & -0.06 & 0.01   & -0.06 & 0.01
         & -0.14 & 0.01 & -0.05 & 0.01
         & -0.21 & 0.01   & -0.19 & 0.01   & -0.14 & 0.01   & -0.08 & 0.01 \\
    \bottomrule
  \end{tabular}
  \label{tab:slope}
\end{table*}

As mentioned before, physically, velocity divergence measures the local compressibility of the fluid, which would favor diffusive shock acceleration (DSA). However, in the highly turbulent and fragmented environment of the simulations analyzed here, well-defined planar shocks are absent. As a result, the link between compression and acceleration weakens, making the shocks less effective. 
These results strengthen the hypothesis that the observed acceleration in our models is not dominated by classical shock processes (such as the DSA), but by a combination of compressive and stochastic mechanisms driven by magnetized turbulence.

Tables 1 and 2 summarize the Pearson correlation coefficients (measuring the strength of statistical correlation) and the regression index obtained from log–log linear regressions between the particle acceleration rate and the physical property of the plasma, respectively.

\section{Discussion}

The correlations between the particle acceleration rate and plasma quantities revealed consistent and relevant patterns for understanding particle acceleration mechanisms in massive binary systems with colliding stellar winds. Among the four analyzed fields — magnetic field intensity ($|\mathbf{B}|$), vorticity ($|\nabla \times \mathbf{v}|$), current density ($|\mathbf{J}|$), and velocity divergence ($\nabla \cdot \mathbf{v}$) — the magnetic field stood out as the variable with the highest average correlation with the acceleration rate. This superlinear increase in acceleration with $|\mathbf{B}|$ suggests that stronger magnetic fields not only confine particles more effectively but also enhance their energy gain rate. These findings support the idea that the magnetic field acts as a key regulator of the acceleration process, as also indicated in earlier more generic works \cite[e.g.][]{kowal2011, Caprioli2014, Marcowith2016}.

Vorticity, in turn, showed correlations very similar to those of the magnetic field (values between 0.46 and 0.50), indicating that turbulence is also strongly associated with enhanced particle acceleration. This may points toward a magnetic turbulence dominated acceleration process, which is typical of turbulent environments where the magnetic field varies chaotically and particles undergo multiple small interactions with dynamic structures \citep{Beresnyak2011, Brunetti2014, Bykov2019}. This hypothesis is further supported by the fragmented nature of the acceleration regions in the simulations, which limits the contribution of well-defined shocks and favors stochastic processes.

Current density exhibited correlation levels similar to those of vorticity (between 0.44 and 0.51), but with lower slopes (ranging from 0.73 to 1.04). Although $\mathbf{J}$ is associated with regions of intense magnetic field reorganization and potentially with magnetic reconnection, this process is negligible at the lengthscales of typical CWBs. From the generalized Ohm’s law, for $\eta \rightarrow 0$, the $\mathbf{v} \times \mathbf{B}$ term dominates, and the acceleration depends strongly on the turbulent structure of the fields rather than on classical resistive dissipation processes \citep{kulsrud2005, lazarian2020}. Even so, the spatial coincidence between regions of high $\mathbf{J}$ and high acceleration rates may reflect the action of turbulent magnetic reconnection -- an efficient mechanism \cite[see][]{dalpino2005, kowal2011, kowal2012particle, lazarian2020}.

Finally, the velocity divergence showed the lowest correlation values (maximum of 0.33) and fitted slopes close to zero or even negative. This indicates that regions of fluid compression or expansion have little direct impact on particle energy gain in the regime studied. These results suggest that acceleration is not dominated by local compressive processes, such as classical shocks, reinforcing the turbulence- and diffusion-driven scenario already outlined. The lack of a significant positive slope also helps to rule out the presence of adiabatic compression acceleration, as occurs in slow, non-relativistic shocks described by the classical DSA model \citep[see, e.g.,][]{Drury1983, Blandford1987}.

In summary, the results of this analysis indicate that velocity divergence has a limited influence on particle acceleration, especially when compared to the other quantities studied (magnetic field, vorticity, and current density). This conclusion aligns with theoretical expectations for environments where turbulence dominates the plasma dynamics, masking the clear signature of classical compressive shocks.

In terms of timescales for acceleration up to very high energies, and the consequential observational detectability, our simulations confirm that ions can reach energies as high as hundreds of TeVs in timescales of order of tens of hours to few days. With such energies, gamma rays with energies in the range of GeV up to tens of TeVs are expected to be observed. In the case of $\eta$~Car, these timescales are small compared to the dynamical timescales of the binary systems, and a rise in gamma rays should be detected as the binaries approach the periastron, with a decrease in gamma ray emission afterwards. This behavior has been confirmed in $\eta$~Car \citep{tavani2009detection, marti2021}. This is a great step-forward in understanding the distribution of relativistic particles emerging from CWBs, compared to previous works that rely on steady-state solutions of the acceleration-diffusion Fokker-Planck equation \citep{Reitberger2014,grimaldo19,pittard20}.

\section{Conclusions}

In this work, we have investigated the process of particle acceleration in colliding-wind binaries (CWBs). We present high resolution 3-D MHD numerical simulations of CWBs, in which a large number of particles are inserted and have their trajectories integrated. From these, we are able to study the particle energy distribution and the correlation of acceleration rates with local plasma properties. 
The main findings of this work can be summarized as follows:

\begin{enumerate}

\item Initially thermal (iron) ions evolve into a relativistic power-law distribution within timescales of hours to days, reaching kinetic energies of several hundred TeV, consistent with steady-state populations over orbital timescales.

\item The highest acceleration rates occur in regions of strong magnetic field and high vorticity, indicating that turbulence-mediated magnetic processes dominate over classical diffusive shock acceleration (DSA).

\item No significant correlation is found between acceleration rate and velocity divergence, suggesting a minor role for shock-driven first-order processes.

\item Current density correlates moderately with acceleration but is likely an indirect tracer of magnetic field complexity rather than a direct driver.

\item Particle energy distributions exhibit a broken power-law: the knee position depends on magnetization, consistent with the Hillas confinement criterion, while the mid-energy slope is magnetization-independent.

\item Cooling enhances turbulence and magnetic field complexity, increasing acceleration efficiency compared to adiabatic shocks.

\end{enumerate}

Overall, our results support a scenario where turbulent, magnetically dominated acceleration is the primary mechanism in CWBs. Future work will address the conversion of these relativistic particles into $\gamma$-rays via distinct radiative processes, assessing their detectability with current and upcoming instruments (e.g. LACT and CTAO).

\section*{Acknowledgements}

The authors acknowledge support from FAPESP (grants 2013/10559-5,
2021/02120-0, 2021/06502-4, and 2022/03972-2) and CAPES. DFG also thanks Conselho Nacional de Desenvolvimento Científico e Tecnológico (CNPq) for grant 304574/2024-4. The simulations presented in this work were performed using the clusters of the Group of Theoretical Astrophysics at EACH-USP (Hydra HPC), which was acquired with support from FAPESP (grants 2013/04073-2 and 2022/03972-2).




\bibliographystyle{mnras}
\bibliography{ms} 

\begin{thebibliography}{}
\makeatletter
\relax
\def\mn@urlcharsother{\let\do\@makeother \do\$\do\&\do\#\do\^\do\_\do\%\do\~}
\def\mn@doi{\begingroup\mn@urlcharsother \@ifnextchar [ {\mn@doi@} {\mn@doi@[]}}
\def\mn@doi@[#1]#2{\def\@tempa{#1}\ifx\@tempa\@empty \href {http://dx.doi.org/#2} {doi:#2}\else \href {http://dx.doi.org/#2} {#1}\fi \endgroup}
\def\mn@eprint#1#2{\mn@eprint@#1:#2::\@nil}
\def\mn@eprint@arXiv#1{\href {http://arxiv.org/abs/#1} {{\tt arXiv:#1}}}
\def\mn@eprint@dblp#1{\href {http://dblp.uni-trier.de/rec/bibtex/#1.xml} {dblp:#1}}
\def\mn@eprint@#1:#2:#3:#4\@nil{\def\@tempa {#1}\def\@tempb {#2}\def\@tempc {#3}\ifx \@tempc \@empty \let \@tempc \@tempb \let \@tempb \@tempa \fi \ifx \@tempb \@empty \def\@tempb {arXiv}\fi \@ifundefined {mn@eprint@\@tempb}{\@tempb:\@tempc}{\expandafter \expandafter \csname mn@eprint@\@tempb\endcsname \expandafter{\@tempc}}}

\bibitem[\protect\citeauthoryear{{Abraham} \& {Falceta-Gon{\c{c}}alves}}{{Abraham} \& {Falceta-Gon{\c{c}}alves}}{2010}]{abraham2010}
{Abraham} Z.,  {Falceta-Gon{\c{c}}alves} D.,  2010, \mn@doi [\mnras] {10.1111/j.1365-2966.2009.15692.x}, \href {https://ui.adsabs.harvard.edu/abs/2010MNRAS.401..687A} {401, 687}

\bibitem[\protect\citeauthoryear{{Abraham}, {Falceta-Gon{\c{c}}alves}, {Dominici}, {Caproni}  \& {Jatenco-Pereira}}{{Abraham} et~al.}{2005}]{abraham2005}
{Abraham} Z.,  {Falceta-Gon{\c{c}}alves} D.,  {Dominici} T.,  {Caproni} A.,   {Jatenco-Pereira} V.,  2005, \mn@doi [\mnras] {10.1111/j.1365-2966.2005.09616.x}, \href {https://ui.adsabs.harvard.edu/abs/2005MNRAS.364..922A} {364, 922}

\bibitem[\protect\citeauthoryear{{Astiasarain}, {Tibaldo}, {Martin}, {Kn{\"o}dlseder}  \& {Remy}}{{Astiasarain} et~al.}{2023}]{cygnusx23}
{Astiasarain} X.,  {Tibaldo} L.,  {Martin} P.,  {Kn{\"o}dlseder} J.,   {Remy} Q.,  2023, \mn@doi [\aap] {10.1051/0004-6361/202245573}, \href {https://ui.adsabs.harvard.edu/abs/2023A&A...671A..47A} {671, A47}

\bibitem[\protect\citeauthoryear{{Bell}}{{Bell}}{1978}]{bell1978}
{Bell} A.~R.,  1978, \mn@doi [\mnras] {10.1093/mnras/182.2.147}, \href {https://ui.adsabs.harvard.edu/abs/1978MNRAS.182..147B} {182, 147}

\bibitem[\protect\citeauthoryear{{Benaglia}, {De Becker}, {Ishwara-Chandra}, {Intema}  \& {Isequilla}}{{Benaglia} et~al.}{2020}]{benaglia2020}
{Benaglia} P.,  {De Becker} M.,  {Ishwara-Chandra} C.~H.,  {Intema} H.~T.,   {Isequilla} N.~L.,  2020, \mn@doi [\pasa] {10.1017/pasa.2020.21}, \href {https://ui.adsabs.harvard.edu/abs/2020PASA...37...30B} {37, e030}

\bibitem[\protect\citeauthoryear{{Benaglia}, {del Palacio}, {Saponara}, {Blanco}, {De Becker}  \& {Marcote}}{{Benaglia} et~al.}{2025}]{benaglia2025}
{Benaglia} P.,  {del Palacio} S.,  {Saponara} J.,  {Blanco} A.~B.,  {De Becker} M.,   {Marcote} B.,  2025, \mn@doi [\aap] {10.1051/0004-6361/202453422}, \href {https://ui.adsabs.harvard.edu/abs/2025A&A...698A..23B} {698, A23}

\bibitem[\protect\citeauthoryear{Beresnyak \& Lazarian}{Beresnyak \& Lazarian}{2011}]{Beresnyak2011}
Beresnyak A.,  Lazarian A.,  2011, The Astrophysical Journal Letters, 729, L29

\bibitem[\protect\citeauthoryear{Blandford \& Eichler}{Blandford \& Eichler}{1987}]{Blandford1987}
Blandford R.,  Eichler D.,  1987, Physics Reports, 154, 1

\bibitem[\protect\citeauthoryear{{Blandford} \& {Ostriker}}{{Blandford} \& {Ostriker}}{1978}]{blandford1978}
{Blandford} R.~D.,  {Ostriker} J.~P.,  1978, \mn@doi [\apjl] {10.1086/182658}, \href {https://ui.adsabs.harvard.edu/abs/1978ApJ...221L..29B} {221, L29}

\bibitem[\protect\citeauthoryear{Brunetti \& Lazarian}{Brunetti \& Lazarian}{2014}]{Brunetti2014}
Brunetti G.,  Lazarian A.,  2014, Monthly Notices of the Royal Astronomical Society, 443, 3564

\bibitem[\protect\citeauthoryear{Bykov, Ellison, Marcowith, Osipov  \& Petruk}{Bykov et~al.}{2019}]{Bykov2019}
Bykov A.~M.,  Ellison D.~C.,  Marcowith A.,  Osipov S.~M.,   Petruk O.,  2019, Space Science Reviews, 215, 14

\bibitem[\protect\citeauthoryear{{Cao} et~al.,}{{Cao} et~al.}{2021}]{llhaso21}
{Cao} Z.,  et~al., 2021, \mn@doi [\nat] {10.1038/s41586-021-03498-z}, \href {https://ui.adsabs.harvard.edu/abs/2021Natur.594...33C} {594, 33}

\bibitem[\protect\citeauthoryear{{Cao} et~al.,}{{Cao} et~al.}{2024}]{llhaso24}
{Cao} Z.,  et~al., 2024, \mn@doi [Science Bulletin] {10.1016/j.scib.2024.07.017}, \href {https://ui.adsabs.harvard.edu/abs/2024SciBu..69.2833C} {69, 2833}

\bibitem[\protect\citeauthoryear{{Cappa}, {Goss}  \& {van der Hucht}}{{Cappa} et~al.}{2004}]{cappa2004}
{Cappa} C.,  {Goss} W.~M.,   {van der Hucht} K.~A.,  2004, \mn@doi [\aj] {10.1086/383286}, \href {https://ui.adsabs.harvard.edu/abs/2004AJ....127.2885C} {127, 2885}

\bibitem[\protect\citeauthoryear{Caprioli}{Caprioli}{2024}]{caprioli2024particle}
Caprioli D.,  2024, in , Foundations of Cosmic Ray Astrophysics.
IOS Press, pp 143--181

\bibitem[\protect\citeauthoryear{Caprioli \& Spitkovsky}{Caprioli \& Spitkovsky}{2014}]{Caprioli2014}
Caprioli D.,  Spitkovsky A.,  2014, The Astrophysical Journal, 783, 91

\bibitem[\protect\citeauthoryear{{De Becker}}{{De Becker}}{2007}]{becker2007}
{De Becker} M.,  2007, \mn@doi [\aapr] {10.1007/s00159-007-0005-2}, \href {https://ui.adsabs.harvard.edu/abs/2007A&ARv..14..171D} {14, 171}

\bibitem[\protect\citeauthoryear{De~Becker \& Raucq}{De~Becker \& Raucq}{2013}]{de2013catalogue}
De~Becker M.,  Raucq F.,  2013, Astronomy \& Astrophysics, 558, A28

\bibitem[\protect\citeauthoryear{{De Becker}, {Benaglia}, {Romero}  \& {Peri}}{{De Becker} et~al.}{2017}]{becker2017}
{De Becker} M.,  {Benaglia} P.,  {Romero} G.~E.,   {Peri} C.~S.,  2017, \mn@doi [\aap] {10.1051/0004-6361/201629110}, \href {https://ui.adsabs.harvard.edu/abs/2017A&A...600A..47D} {600, A47}

\bibitem[\protect\citeauthoryear{{Dougherty}, {Beasley}, {Claussen}, {Zauderer}  \& {Bolingbroke}}{{Dougherty} et~al.}{2005}]{doug2005}
{Dougherty} S.~M.,  {Beasley} A.~J.,  {Claussen} M.~J.,  {Zauderer} B.~A.,   {Bolingbroke} N.~J.,  2005, \mn@doi [\apj] {10.1086/428494}, \href {https://ui.adsabs.harvard.edu/abs/2005ApJ...623..447D} {623, 447}

\bibitem[\protect\citeauthoryear{Drury}{Drury}{1983}]{Drury1983}
Drury L.~O.,  1983, Reports on Progress in Physics, 46, 973

\bibitem[\protect\citeauthoryear{Eichler \& Usov}{Eichler \& Usov}{1993}]{eichler1993particle}
Eichler D.,  Usov V.,  1993, Astrophysical Journal, Part 1 (ISSN 0004-637X), vol. 402, no. 1, p. 271-279., 402, 271

\bibitem[\protect\citeauthoryear{{Falceta-Goncalves}}{{Falceta-Goncalves}}{2015}]{falceta15}
{Falceta-Goncalves} D.,  2015, in {Hamann} W.-R.,  {Sander} A.,   {Todt} H.,  eds, Wolf-Rayet Stars. pp 289--292 (\mn@eprint {arXiv} {1510.06106}), \mn@doi{10.48550/arXiv.1510.06106}

\bibitem[\protect\citeauthoryear{{Falceta-Gon{\c{c}}alves} \& {Abraham}}{{Falceta-Gon{\c{c}}alves} \& {Abraham}}{2012}]{falceta2012}
{Falceta-Gon{\c{c}}alves} D.,  {Abraham} Z.,  2012, \mn@doi [\mnras] {10.1111/j.1365-2966.2012.20978.x}, \href {https://ui.adsabs.harvard.edu/abs/2012MNRAS.423.1562F} {423, 1562}

\bibitem[\protect\citeauthoryear{{Falceta-Gon{\c{c}}alves}, {Jatenco-Pereira}  \& {Abraham}}{{Falceta-Gon{\c{c}}alves} et~al.}{2005}]{falceta2005}
{Falceta-Gon{\c{c}}alves} D.,  {Jatenco-Pereira} V.,   {Abraham} Z.,  2005, \mn@doi [\mnras] {10.1111/j.1365-2966.2005.08682.x}, \href {https://ui.adsabs.harvard.edu/abs/2005MNRAS.357..895F} {357, 895}

\bibitem[\protect\citeauthoryear{{Falceta-Gon{\c{c}}alves}, {Kowal}, {Falgarone}  \& {Chian}}{{Falceta-Gon{\c{c}}alves} et~al.}{2014}]{falceta2014b}
{Falceta-Gon{\c{c}}alves} D.,  {Kowal} G.,  {Falgarone} E.,   {Chian} A.~C.~L.,  2014, \mn@doi [Nonlinear Processes in Geophysics] {10.5194/npg-21-587-2014}, \href {https://ui.adsabs.harvard.edu/abs/2014NPGeo..21..587F} {21, 587}

\bibitem[\protect\citeauthoryear{{Fermi}}{{Fermi}}{1949}]{fermi1949}
{Fermi} E.,  1949, \mn@doi [Physical Review] {10.1103/PhysRev.75.1169}, \href {https://ui.adsabs.harvard.edu/abs/1949PhRv...75.1169F} {75, 1169}

\bibitem[\protect\citeauthoryear{{Ginzburg} \& {Syrovatsky}}{{Ginzburg} \& {Syrovatsky}}{1961}]{ginzburg1961}
{Ginzburg} V.~L.,  {Syrovatsky} S.~I.,  1961, \mn@doi [Progress of Theoretical Physics Supplement] {10.1143/PTPS.20.1}, \href {https://ui.adsabs.harvard.edu/abs/1961PThPS..20....1G} {20, 1}

\bibitem[\protect\citeauthoryear{Gottlieb, Ketcheson  \& Shu}{Gottlieb et~al.}{2011}]{gottlieb2011strong}
Gottlieb S.,  Ketcheson D.,   Shu C.-W.,  2011, Strong stability preserving Runge-Kutta and multistep time discretizations.
World Scientific

\bibitem[\protect\citeauthoryear{{Grimaldo}, {Reimer}, {Kissmann}, {Niederwanger}  \& {Reitberger}}{{Grimaldo} et~al.}{2019}]{grimaldo19}
{Grimaldo} E.,  {Reimer} A.,  {Kissmann} R.,  {Niederwanger} F.,   {Reitberger} K.,  2019, \mn@doi [\apj] {10.3847/1538-4357/aaf6ee}, \href {https://ui.adsabs.harvard.edu/abs/2019ApJ...871...55G} {871, 55}

\bibitem[\protect\citeauthoryear{{Hairer} \& {Pavliotis}}{{Hairer} \& {Pavliotis}}{2008}]{hairer2008}
{Hairer} M.,  {Pavliotis} G.~A.,  2008, \mn@doi [Journal of Statistical Physics] {10.1007/s10955-008-9493-3}, \href {https://ui.adsabs.harvard.edu/abs/2008JSP...131..175H} {131, 175}

\bibitem[\protect\citeauthoryear{Hamaguchi et~al.,}{Hamaguchi et~al.}{2018}]{hamaguchi2018non}
Hamaguchi K.,  et~al., 2018, Nature Astronomy, 2, 731

\bibitem[\protect\citeauthoryear{{Jardine}, {Allen}  \& {Pollock}}{{Jardine} et~al.}{1996}]{jardine1996}
{Jardine} M.,  {Allen} H.~R.,   {Pollock} A.~M.~T.,  1996, \aap, \href {https://ui.adsabs.harvard.edu/abs/1996A&A...314..594J} {314, 594}

\bibitem[\protect\citeauthoryear{{Jokipii}}{{Jokipii}}{1977}]{jokipii1977}
{Jokipii} J.~R.,  1977, in International Cosmic Ray Conference. p.~429

\bibitem[\protect\citeauthoryear{{Kowal} \& {Falceta-Gon{\c{c}}alves}}{{Kowal} \& {Falceta-Gon{\c{c}}alves}}{2021}]{kowal.2021}
{Kowal} G.,  {Falceta-Gon{\c{c}}alves} D.~A.,  2021, \mn@doi [Frontiers in Astronomy and Space Sciences] {10.3389/fspas.2021.667805}, \href {https://ui.adsabs.harvard.edu/abs/2021FrASS...8...75K} {8, 75}

\bibitem[\protect\citeauthoryear{{Kowal}, {de Gouveia Dal Pino}  \& {Lazarian}}{{Kowal} et~al.}{2011}]{kowal2011}
{Kowal} G.,  {de Gouveia Dal Pino} E.~M.,   {Lazarian} A.,  2011, \mn@doi [\apj] {10.1088/0004-637X/735/2/102}, \href {https://ui.adsabs.harvard.edu/abs/2011ApJ...735..102K} {735, 102}

\bibitem[\protect\citeauthoryear{{Kowal}, {de Gouveia Dal Pino}  \& {Lazarian}}{{Kowal} et~al.}{2012}]{kowal2012particle}
{Kowal} G.,  {de Gouveia Dal Pino} E.~M.,   {Lazarian} A.,  2012, \mn@doi [\prl] {10.1103/PhysRevLett.108.241102}, \href {https://ui.adsabs.harvard.edu/abs/2012PhRvL.108x1102K} {108, 241102}

\bibitem[\protect\citeauthoryear{{Kulsrud}}{{Kulsrud}}{2005}]{kulsrud2005}
{Kulsrud} R.~M.,  2005, {Plasma Physics for Astrophysics}.
{Princeton Univ. Press}

\bibitem[\protect\citeauthoryear{{Lazarian} \& {Vishniac}}{{Lazarian} \& {Vishniac}}{1999}]{Lazarian1999}
{Lazarian} A.,  {Vishniac} E.~T.,  1999, \mn@doi [\apj] {10.1086/307233}, \href {https://ui.adsabs.harvard.edu/abs/1999ApJ...517..700L} {517, 700}

\bibitem[\protect\citeauthoryear{{Lazarian}, {Eyink}, {Jafari}, {Kowal}, {Li}, {Xu}  \& {Vishniac}}{{Lazarian} et~al.}{2020}]{lazarian2020}
{Lazarian} A.,  {Eyink} G.~L.,  {Jafari} A.,  {Kowal} G.,  {Li} H.,  {Xu} S.,   {Vishniac} E.~T.,  2020, \mn@doi [Physics of Plasmas] {10.1063/1.5110603}, \href {https://ui.adsabs.harvard.edu/abs/2020PhPl...27a2305L} {27, 012305}

\bibitem[\protect\citeauthoryear{Marcowith et~al.,}{Marcowith et~al.}{2016}]{Marcowith2016}
Marcowith A.,  et~al., 2016, Reports on Progress in Physics, 79, 046901

\bibitem[\protect\citeauthoryear{{Mart{\'\i}-Devesa} \& {Reimer}}{{Mart{\'\i}-Devesa} \& {Reimer}}{2021}]{marti2021}
{Mart{\'\i}-Devesa} G.,  {Reimer} O.,  2021, \mn@doi [\aap] {10.1051/0004-6361/202140451}, \href {https://ui.adsabs.harvard.edu/abs/2021A&A...654A..44M} {654, A44}

\bibitem[\protect\citeauthoryear{{Mart{\'\i}-Devesa}, {Reimer}, {Li}  \& {Torres}}{{Mart{\'\i}-Devesa} et~al.}{2020}]{marti2020}
{Mart{\'\i}-Devesa} G.,  {Reimer} O.,  {Li} J.,   {Torres} D.~F.,  2020, \mn@doi [\aap] {10.1051/0004-6361/202037462}, \href {https://ui.adsabs.harvard.edu/abs/2020A&A...635A.141M} {635, A141}

\bibitem[\protect\citeauthoryear{Mignone}{Mignone}{2007}]{mignone2007simple}
Mignone A.,  2007, Journal of Computational Physics, 225, 1427

\bibitem[\protect\citeauthoryear{{Monnier}, {Greenhill}, {Tuthill}  \& {Danchi}}{{Monnier} et~al.}{2002}]{monnier2002}
{Monnier} J.~D.,  {Greenhill} L.~J.,  {Tuthill} P.~G.,   {Danchi} W.~C.,  2002, \mn@doi [\apj] {10.1086/337961}, \href {https://ui.adsabs.harvard.edu/abs/2002ApJ...566..399M} {566, 399}

\bibitem[\protect\citeauthoryear{{Morlino}, {Blasi}, {Peretti}  \& {Cristofari}}{{Morlino} et~al.}{2021}]{morlino21}
{Morlino} G.,  {Blasi} P.,  {Peretti} E.,   {Cristofari} P.,  2021, \mn@doi [\mnras] {10.1093/mnras/stab690}, \href {https://ui.adsabs.harvard.edu/abs/2021MNRAS.504.6096M} {504, 6096}

\bibitem[\protect\citeauthoryear{Pittard \& Dougherty}{Pittard \& Dougherty}{2006}]{pittard2006radio}
Pittard J.,  Dougherty S.,  2006, Monthly Notices of the Royal Astronomical Society, 372, 801

\bibitem[\protect\citeauthoryear{{Pittard}, {Vila}  \& {Romero}}{{Pittard} et~al.}{2020}]{pittard20}
{Pittard} J.~M.,  {Vila} G.~S.,   {Romero} G.~E.,  2020, \mn@doi [\mnras] {10.1093/mnras/staa1099}, \href {https://ui.adsabs.harvard.edu/abs/2020MNRAS.495.2205P} {495, 2205}

\bibitem[\protect\citeauthoryear{{Rauw}}{{Rauw}}{2025}]{rauw2025}
{Rauw} G.,  2025, \mn@doi [Contributions of the Astronomical Observatory Skalnate Pleso] {10.31577/caosp.2025.55.3.75}, \href {https://ui.adsabs.harvard.edu/abs/2025CoSka..55c..75R} {55, 75}

\bibitem[\protect\citeauthoryear{{Rauw} \& {Naz{\'e}}}{{Rauw} \& {Naz{\'e}}}{2016}]{rauw2016}
{Rauw} G.,  {Naz{\'e}} Y.,  2016, \mn@doi [Advances in Space Research] {10.1016/j.asr.2015.09.026}, \href {https://ui.adsabs.harvard.edu/abs/2016AdSpR..58..761R} {58, 761}

\bibitem[\protect\citeauthoryear{{Reitberger}, {Kissmann}, {Reimer}  \& {Reimer}}{{Reitberger} et~al.}{2014}]{Reitberger2014}
{Reitberger} K.,  {Kissmann} R.,  {Reimer} A.,   {Reimer} O.,  2014, \mn@doi [\apj] {10.1088/0004-637X/789/1/87}, \href {https://ui.adsabs.harvard.edu/abs/2014ApJ...789...87R} {789, 87}

\bibitem[\protect\citeauthoryear{{Reyes}}{{Reyes}}{2019}]{2019PhDT........51R}
{Reyes} A.~C.,  2019, PhD thesis, University of California, Santa Cruz

\bibitem[\protect\citeauthoryear{{Reyes}, {Lee}, {Graziani}  \& {Tzeferacos}}{{Reyes} et~al.}{2016}]{2016arXiv161108084R}
{Reyes} A.,  {Lee} D.,  {Graziani} C.,   {Tzeferacos} P.,  2016, \mn@doi [arXiv e-prints] {10.48550/arXiv.1611.08084}, \href {https://ui.adsabs.harvard.edu/abs/2016arXiv161108084R} {p. arXiv:1611.08084}

\bibitem[\protect\citeauthoryear{{Rocha da Silva}, {Falceta-Gon{\c{c}}alves}, {Kowal}  \& {de Gouveia Dal Pino}}{{Rocha da Silva} et~al.}{2015}]{fg2015}
{Rocha da Silva} G.,  {Falceta-Gon{\c{c}}alves} D.,  {Kowal} G.,   {de Gouveia Dal Pino} E.~M.,  2015, \mn@doi [\mnras] {10.1093/mnras/stu2104}, \href {https://ui.adsabs.harvard.edu/abs/2015MNRAS.446..104R} {446, 104}

\bibitem[\protect\citeauthoryear{{Saha}, {Tej}, {del Palacio}, {De Becker}, {Benaglia}, {Ishwara-Chandra}  \& {Prajapati}}{{Saha} et~al.}{2023}]{saha2023}
{Saha} A.,  {Tej} A.,  {del Palacio} S.,  {De Becker} M.,  {Benaglia} P.,  {Ishwara-Chandra} C.~H.,   {Prajapati} P.,  2023, \mn@doi [\mnras] {10.1093/mnras/stad2758}, \href {https://ui.adsabs.harvard.edu/abs/2023MNRAS.526..750S} {526, 750}

\bibitem[\protect\citeauthoryear{{Skinner}, {Itoh}, {Nagase}  \& {Zhekov}}{{Skinner} et~al.}{1999}]{skinner1999}
{Skinner} S.~L.,  {Itoh} M.,  {Nagase} F.,   {Zhekov} S.~A.,  1999, \mn@doi [\apj] {10.1086/307809}, \href {https://ui.adsabs.harvard.edu/abs/1999ApJ...524..394S} {524, 394}

\bibitem[\protect\citeauthoryear{{Stevens}, {Blondin}  \& {Pollock}}{{Stevens} et~al.}{1992}]{stevens1992}
{Stevens} I.~R.,  {Blondin} J.~M.,   {Pollock} A.~M.~T.,  1992, \mn@doi [\apj] {10.1086/171013}, \href {https://ui.adsabs.harvard.edu/abs/1992ApJ...386..265S} {386, 265}

\bibitem[\protect\citeauthoryear{Tavani et~al.,}{Tavani et~al.}{2009}]{tavani2009detection}
Tavani M.,  et~al., 2009, The Astrophysical Journal, 698, L142

\bibitem[\protect\citeauthoryear{{Tsuji} et~al.,}{{Tsuji} et~al.}{2025}]{tsuji25}
{Tsuji} N.,  et~al., 2025, \mn@doi [\apj] {10.3847/1538-4357/adb7df}, \href {https://ui.adsabs.harvard.edu/abs/2025ApJ...983...22T} {983, 22}

\bibitem[\protect\citeauthoryear{Werner, Reimer, Reimer  \& Egberts}{Werner et~al.}{2013}]{werner2013fermi}
Werner M.,  Reimer O.,  Reimer A.,   Egberts K.,  2013, Astronomy \& Astrophysics, 555, A102

\bibitem[\protect\citeauthoryear{{White} \& {Becker}}{{White} \& {Becker}}{1995}]{white1995}
{White} R.~L.,  {Becker} R.~H.,  1995, \mn@doi [\apj] {10.1086/176224}, \href {https://ui.adsabs.harvard.edu/abs/1995ApJ...451..352W} {451, 352}

\bibitem[\protect\citeauthoryear{Wiedemann}{Wiedemann}{2015}]{wiedemann2015particle}
Wiedemann H.,  2015, Particle accelerator physics.
Springer Nature

\bibitem[\protect\citeauthoryear{{de Gouveia Dal Pino}, {Falceta-Gon{\c{c}}alves}, {Gallagher}, {Melioli}, {D'Ercole}  \& {Brighenti}}{{de Gouveia Dal Pino} et~al.}{2010}]{pino2009role}
{de Gouveia Dal Pino} E.~M.,  {Falceta-Gon{\c{c}}alves} D.,  {Gallagher} J.~S.,  {Melioli} C.,  {D'Ercole} A.,   {Brighenti} F.,  2010, \mn@doi [Highlights of Astronomy] {10.1017/S1743921310010240}, \href {https://ui.adsabs.harvard.edu/abs/2010HiA....15..452D} {15, 452}

\bibitem[\protect\citeauthoryear{{de Gouveia dal Pino} \& {Lazarian}}{{de Gouveia dal Pino} \& {Lazarian}}{2005}]{dalpino2005}
{de Gouveia dal Pino} E.~M.,  {Lazarian} A.,  2005, \mn@doi [\aap] {10.1051/0004-6361:20042590}, \href {https://ui.adsabs.harvard.edu/abs/2005A&A...441..845D} {441, 845}

\bibitem[\protect\citeauthoryear{{van Leer}}{{van Leer}}{1979}]{vl1979}
{van Leer} B.,  1979, \mn@doi [Journal of Computational Physics] {10.1016/0021-9991(79)90145-1}, \href {https://ui.adsabs.harvard.edu/abs/1979JCoPh..32..101V} {32, 101}

\makeatother
\end{thebibliography}




\appendix


\bsp	
\label{lastpage}
\end{document}